\documentclass[
reprint,
superscriptaddress,
amsmath,
amssymb,
aps,
longbibliography,
pra, 
showpacs,
floatfix
]{revtex4-1}

\usepackage{graphicx}
\usepackage{color}

\newcommand{\units}[2]{\ensuremath{#1\,\text{#2}}}
\newcommand{\ang}{\ensuremath{\text{\AA}}}

\begin{document}

\title{Sum-frequency ionic Raman scattering}

\author{Dominik~M.\ Juraschek}
\affiliation{Materials Theory, ETH Zurich, CH-8093 Z\"{u}rich, Switzerland}
\author{Sebastian~F. Maehrlein}
\affiliation{Department of Physical Chemistry, Fritz Haber Institute of the Max Planck Society, DE-14195 Berlin, Germany}
\affiliation{Department of Chemistry, Columbia University, New York, NY 10027, USA}

\date{\today}

\begin{abstract}
In a recent report sum-frequency excitation of a Raman-active phonon was experimentally demonstrated for the first time. This mechanism is the sibling of impulsive stimulated Raman scattering, in which difference-frequency components of a light field excite a Raman-active mode. Here we propose that \textit{ionic} Raman scattering analogously has a sum-frequency counterpart. We compare the four Raman mechanisms, photonic and ionic difference- and sum-frequency excitation, for three different example materials using a generalized oscillator model for which we calculate the parameters with density functional theory. Sum-frequency ionic Raman scattering completes the toolkit for controlling  materials properties by means of selective excitation of lattice vibrations.
\end{abstract}

\maketitle


\section{Introduction}

Ultrashort electromagnetic pulses are an established tool to control the electronic and structural phases of matter. Intense laser pulses in the terahertz spectral range provide direct access to the excitation of optical phonons and have become practical only during the past decade \cite{juraschek3:2017, Kampfrath2013a, Dhillon2017, Sell2008}. Highly excited optical phonons govern a variety of physical phenomena, such as phase transitions \cite{Rini2007}, induced or enhanced superconductivity \cite{fausti:2011,hu:2014,Mitrano2016}, and control of magnetic order \cite{nova:2017,Maehrlein2018}. It is therefore necessary to understand the fundamental mechanisms that underly the excitation of coherent optical phonons with light. Infrared-active phonons carry an electric dipole moment and can therefore be excited directly by coupling to the electric field component of electromagnetic radiation. For Raman-active phonons, which do not possess an electric dipole moment, another, indirect way has to be taken. 

\begin{figure}[t]
\includegraphics[scale=0.195]{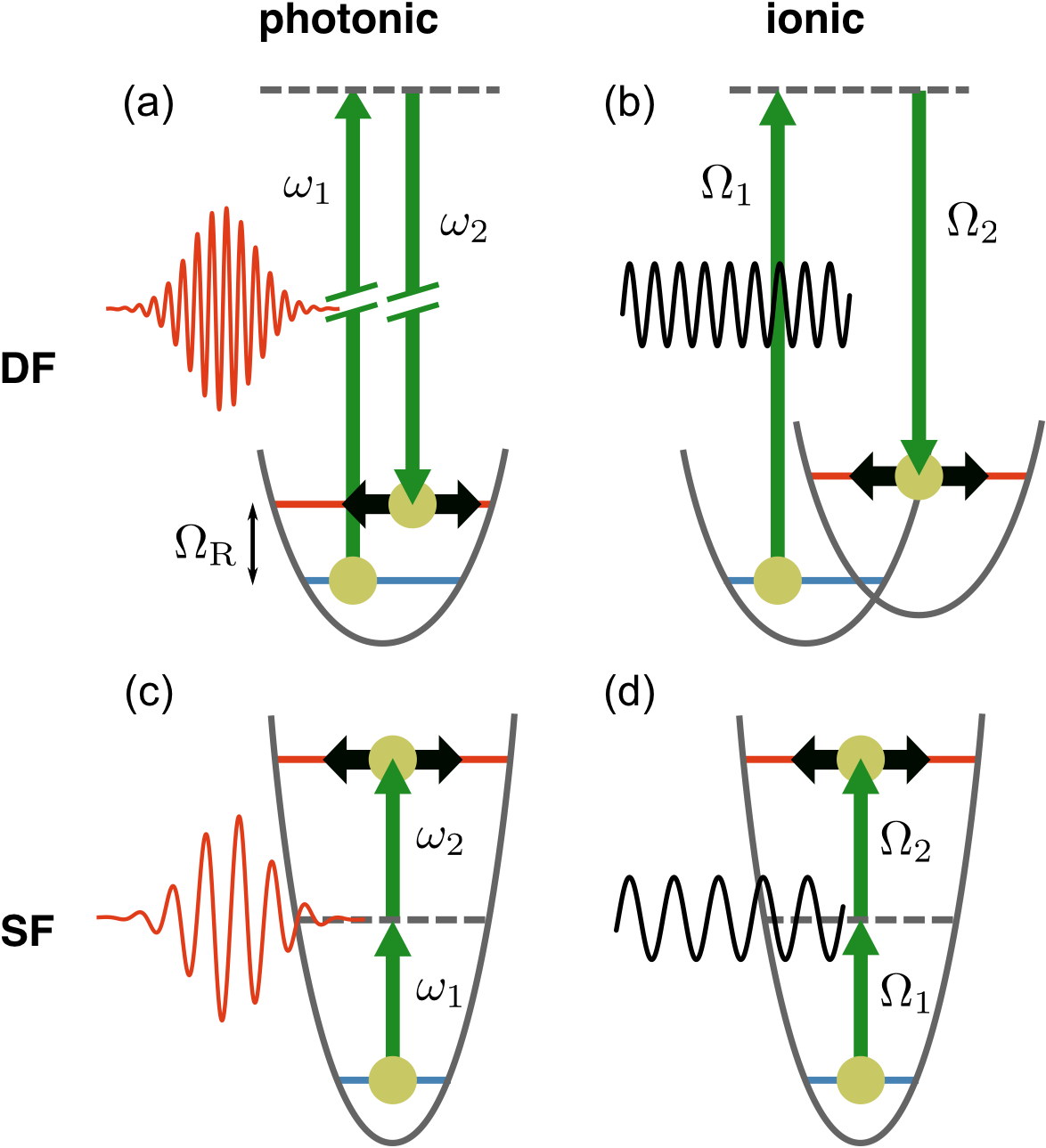}~~
\caption{
\label{fig:mechanisms}
Map of photonic and ionic Raman mechanisms. The difference- and sum-frequency components of a light pulse, $\omega_1\pm\omega_2$, or an infrared-active phonon, $\Omega_\text{1}\pm\Omega_\text{2}$, are resonant with a vibrational transition of a Raman-active phonon, $\Omega_\text{R}$. (a) Impulsive stimulated Raman scattering (ISRS). (b) Conventional ionic Raman scattering (DF-IRS). The lattice potential of the Raman-active mode shifts. (c) Terahertz sum-frequency excitation (THz-SFE). (d) Proposed sum-frequency counterpart of ionic Raman scattering (SF-IRS).
}
\end{figure}

An established route is to disturb the electronic system with an ultrashort light pulse, which then mediates energy to Raman-active phonons via electron-phonon interaction \cite{Dekorsy2000,Forst2008}. For coherent, nonresonant excitation below the band gap \footnote{Under photoexcitation there are a lot more mechanisms that excite Raman-active phonons, such as ``displacive excitation of coherent phonons'' or ``transient depletion field screening''.}, the primary mechanism involved is impulsive stimulated Raman scattering (ISRS), in which a virtual electronic state serves as intermediate energy level for the Raman scattering of the incident light by the phonon, see figure~\ref{fig:mechanisms}(a) \cite{desilvestri:1985,merlin:1997,Stevens2002}. In this case, the difference frequency of two photons from the light pulse is resonant with a vibrational transition of the phonon mode. We will in the following refer to this as a ``photonic'' Raman mechanism.

A second route to exciting Raman-active phonons is via scattering with infrared-active phonons. This was proposed nearly half a century ago as ionic Raman scattering (IRS) and has only been demonstrated within this decade due to the advancement of intense terahertz sources \cite{maradurin:1970,forst:2011}. In ionic Raman scattering, a highly excited infrared-active phonon serves as the intermediate state for Raman scattering, see figure~\ref{fig:mechanisms}(b). This effect is mediated through anharmonic phonon-phonon coupling rather than electron-phonon interaction, and it is less dissipative than its photonic counterpart due to the lower energy of the excitation \cite{subedi:2014,nicoletti:2016}. 

In a recent experiment, a third route has been demonstrated, in which the \units{40}{THz} Raman-active phonon of diamond is excited by a terahertz pulse in a two-photon absorption process, see figure~\ref{fig:mechanisms}(c) \cite{maehrlein:2017}. This is the sum-frequency excitation (SFE) counterpart to ISRS, which combines the possibility to excite phonons in compounds that do not possess infrared-active phonons with the advantage of low-energy excitation by terahertz radiation.

The purpose of this theoretical study is two fold: First, we complete the map of photonic and ionic difference- and sum-frequency Raman mechanisms with the ``missing'' sum-frequency part of ionic Raman scattering (SF-IRS), see figure~\ref{fig:mechanisms}(d). Second, we compare the four mechanisms for three different example materials: Diamond, erbium ferrite (ErFeO$_3$), and bismuth ferrite (BiFeO$_3$). Assuming realistic experimental conditions, we show that sum-frequency excitation, both photonic and ionic, are able to coherently control Raman-active phonons in the electronic ground state in a way that is complementary to previous nonlinear phononics studies.

\section{Theoretical model}

\subsection{General equation of motion for the excitation of phonons}

To investigate the time evolution of a phonon mode we numerically solve its equation of motion:
\begin{equation}\label{eq:eom_evaluated}
\ddot{Q} + \kappa\dot{Q} + \frac{\partial}{\partial Q}V(Q) = \sum\limits_i Z_i E_i + \varepsilon_0 \sum\limits_{ij} R_{ij} E_i E_j,
\end{equation}
where $Q$ is the normal mode coordinate (or amplitude) of the phonon at the Brillouin-zone center in units of $\sqrt{\mu}\ang$, with $\mu$ being the atomic mass unit. $\kappa$ is the phonon linewidth, and $V(Q)$ is the (an)harmonic lattice potential of the phonon. $E_i$ is the electric field component of the exciting electromagnetic pulse, and the indices $i$ and $j$ run over the spatial coordinates. $Z_i = \big(\sum_n \textbf{Z}^\ast_n \textbf{u}_n\big)_i$ is the mode effective charge of the phonon with $\textbf{Z}^\ast_n$ being the Born effective charge tensor and $\textbf{u}_n$ the displacement vector of atom $n$, and the sum runs over all atoms in the unit cell \cite{Gonze1997}. $R_{ij}=V_\text{c}\partial_Q\chi_{ij}$ is the Raman tensor, which is given by the linear electric susceptibility, $\chi_{ij}$, and the volume of the unit cell, $V_\text{c}$. Phonon linewidths lie in the range of $\kappa\approx\Omega/10~\text{to}~\Omega/20$ for the materials that we consider here, where $\Omega$ is the eigenfrequency of the phonon mode \cite{nova:2017,juraschek:2017,maehrlein:2017}. For a detailed derivation, see for example reference~\cite{Dhar1994}. We model the electric field component of a light or terahertz pulse as $E(t) = E_0 \text{exp}(-(t-t_0)^2/(2(\tau/2\sqrt{2\text{ln}2})^2)) \cos(\omega_0 t+\varphi_\text{CEP})$, where $E_0$ is the peak electric field, $\omega_0$ is the center frequency, $\varphi_\text{CEP}$ is the carrier-envelope phase, and $\tau$ is the full width at half maximum duration of the pulse.

\subsection{Equations of motion for photonic and ionic Raman scattering}

For photonic Raman scattering we assume a harmonic lattice potential, $V(Q_\text{R})=\Omega_\text{R}^2Q_\text{R}^2/2$, as anharmonicities are not important for the process. Furthermore in centrosymmetric crystals $Z_i=0$ for Raman-active phonons, and for a linearly polarized laser pulse, we can write equation~\ref{eq:eom_evaluated} as
\begin{equation}\label{eq:eom_photonic_R}
\ddot{Q}_\text{R} + \kappa_\text{R}\dot{Q}_\text{R} + \Omega_\text{R}^2Q_\text{R} = \varepsilon_0 R E^2(t).
\end{equation}
In contrast, ionic Raman scattering is described by a quadratic-linear coupling of an infrared-active with a Raman-active phonon. The anharmonic lattice potential in its simplest form can be expressed as $V(Q_\text{R},Q_\text{IR})=\Omega_\text{R}^2Q_\text{R}^2/2+\Omega_\text{IR}^2Q_\text{IR}^2/2+cQ_\text{IR}^2Q_\text{R}$, where $c$ is the quadratic-linear coupling coefficient given in meV/($\sqrt{\mu}$\ang)$^3$ \cite{subedi:2014}. We have to solve the equations of motion for both phonons, and equation~\ref{eq:eom_evaluated} can be written respectively as
\begin{align}
\ddot{Q}_\text{IR} + \kappa_\text{IR}\dot{Q}_\text{IR} + (\Omega_\text{IR}^2+2cQ_\text{R})Q_\text{IR} & = Z_\text{IR} E(t), \label{eq:eom_ionic_IR} \\
\ddot{Q}_\text{R} + \kappa_\text{R}\dot{Q}_\text{R} + \Omega_\text{R}^2Q_\text{R} & = cQ_\text{IR}^2(t). \label{eq:eom_ionic_R}
\end{align}

The driving force of the Raman-active phonon in photonic Raman scattering is the square of the electric field, $E^2(t)$, see equation~\ref{eq:eom_photonic_R}, while in ionic Raman scattering it is the square of the ``phonon field'' of the infrared-active phonon, $Q^2_\text{IR}(t)$, see equation~\ref{eq:eom_ionic_R}. In addition, the Raman-active mode feedback affects the initially excited infrared-active mode by dynamically renormalizing its frequency as $\Omega^2_\text{IR}\rightarrow\Omega^2_\text{IR}+2cQ_\text{R}$, see equation~\ref{eq:eom_ionic_IR}.

The two photonic processes, ISRS and THz-SFE, can be described by the same equation of motion \ref{eq:eom_photonic_R}. The two mechanisms are only distinguished by the duration of the pulse and its center frequency, $\omega_0$, which is higher than the phonon frequency in ISRS, $\omega_0>\Omega_\text{R}$, and ideally half the phonon frequency in THz-SFE, $\omega_0=\Omega_\text{R}/2$. We can draw an analogy for ionic Raman scattering here, which has so far been always connected to the coupling of a high-frequency infrared-active phonon with a low-frequency Raman-active phonon, $\Omega_\text{IR}>\Omega_\text{R}$, in which difference-frequency components of the phonon field $Q^2_\text{IR}$ are responsible for the excitation of the Raman-active phonon \cite{forst:2011,subedi:2014,subedi:2015,fechner:2016,subedi:2017,Mankowsky_2:2017,Gu2017,Gu_2:2017,Fechner2017,Itin2017}. Here, we will show that this mechanism can be extended to a sum-frequency counterpart that fulfills $\Omega_\text{IR}=\Omega_\text{R}/2$ just analog to the photonic Raman processes \cite{maehrlein:2017}. A summary of the discussion in this section is given in table \ref{tab:mechanismsproperties}, in which also the results for phase sensitivity and impulsiveness from the following sections are shown.

\vspace{-0.1cm}

\subsection{Computational details}

\begin{table*}[t]
\centering
\bgroup
\def\arraystretch{1.3}
\caption{
Summary of properties of the four mechanisms for the excitation of Raman-active phonons in insulators.
}
\begin{tabular}{c | c c c c}
\hline\hline
 & ISRS & DF-IRS & THz-SFE & SF-IRS \\ 
\hline
Type of excitation & photonic & ionic & photonic & ionic \\
Driving force & $E^2(t)$ & $Q^2_\text{IR}(t)$ & $E^2(t)$ & $Q^2_\text{IR}(t)$ \\
Center frequency & $\omega_0 > \Omega_\text{R}$ & $\Omega_\text{IR} > \Omega_\text{R}$ & $\omega_0 = \Omega_\text{R}/2$ & $\Omega_\text{IR} = \Omega_\text{R}/2$ \\
Frequency components & difference & difference & sum & sum \\
CEP sensitive & no & no & yes & yes\\
Impulsive & yes & yes & no & no\\
\hline\hline
\end{tabular}
\label{tab:mechanismsproperties}
\egroup
\end{table*}

We calculated the phonon eigenfrequencies, eigenvectors, and the Raman tensors from first-principles using the density functional theory formalism as implemented in the Vienna ab-initio simulation package (VASP) \cite{kresse:1996,kresse2:1996}, and the frozen-phonon method as implemented in the phonopy package \cite{phonopy}. To calculate the frequency-dependent Raman tensor we followed the scheme of reference~\cite{porezag:1996}. We used the default VASP PAW pseudopotentials for every considered atom and converged the Hellmann-Feynman forces to \units{10^{-5}}{eV/\ang{}} using a plane-wave energy cut-off of \units{950}{eV} and a 9$\times$9$\times$9 $k$-point Monkhorst-Pack mesh \cite{Monkhorst/Pack:1976} to sample the Brillouin zone for diamond and \units{850}{eV}, 6$\times$6$\times$6 for BiFeO$_3$. For the exchange-correlation functional, we chose the PBEsol form of the generalized gradient approximation (GGA) \cite{csonka:2009}. For BiFeO$_3$ we found that an on-site Coulomb interaction of $\units{4}{eV}$ and a Hund's exchange of $\units{1}{eV}$ optimally reproduce both the $G$-type antiferromagnetic ordering and lattice dynamical properties \cite{Ederer/Spaldin:2005,Wei2016}. Our fully relaxed structures with lattice constants $\units{3.55}{\ang}$ for diamond and $\units{3.94}{\ang}$ with pseudocubic angle 90.44$^\circ$ for BiFeO$_3$ fit reasonably well to common experimental values \cite{Sosnowska_et_al:2002,Mildren2013}, as do our calculated phonon frequencies. Our calculated phonon eigenfrequency for the $F_\text{2g}$ mode in diamond is $39.2\,\text{THz}$, though for simplicity we keep referring to it as the ``40\,THz mode''. For the details on ErFeO$_3$, we refer the reader to the computational details of reference~\cite{juraschek:2017}.

\section{Results}

\subsection{THz-SFE versus ISRS in diamond}

\begin{figure*}[t]
\includegraphics[scale=0.31]{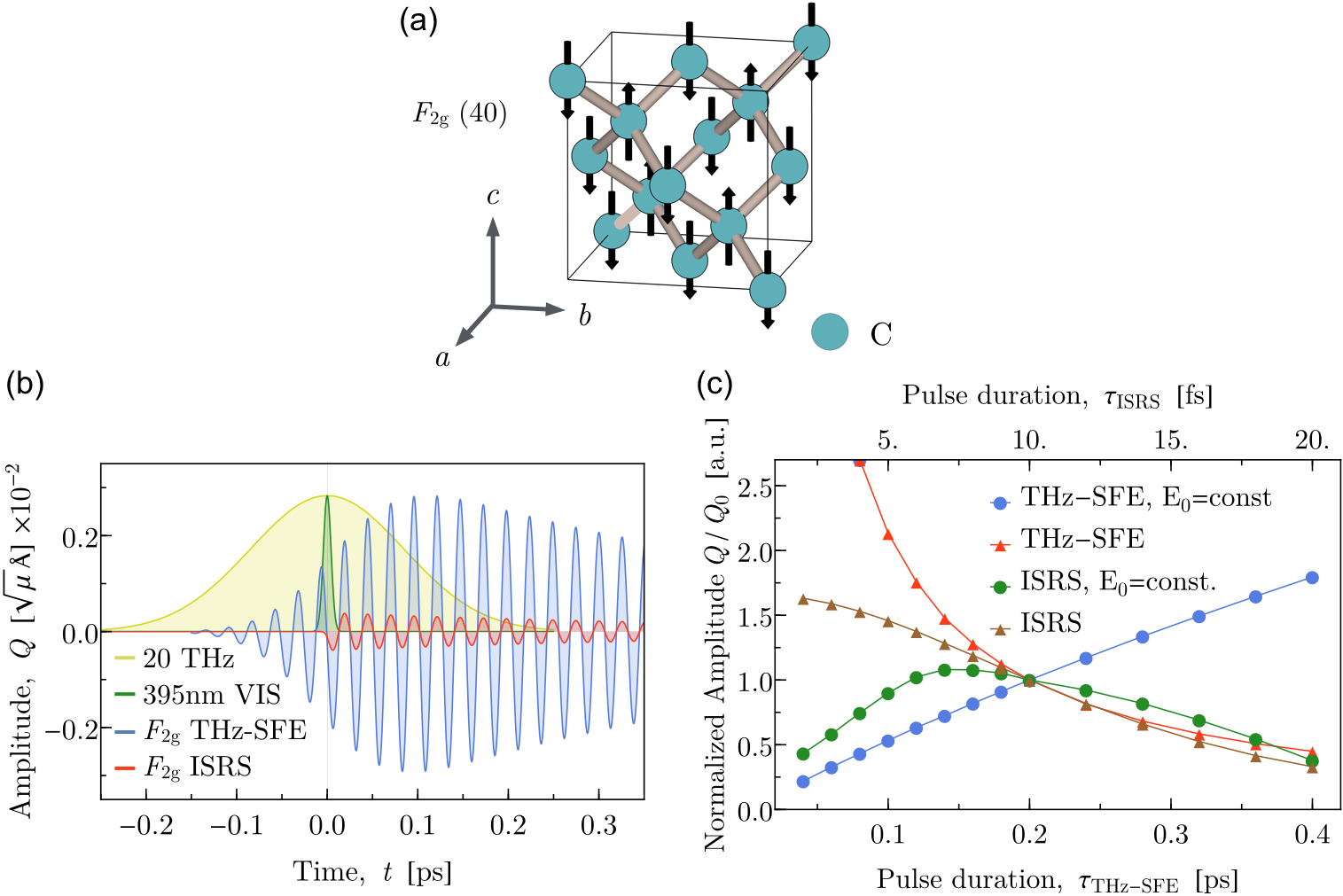}
\caption{
\label{fig:diamond}
(a) Eigenvector of the \units{40}{THz} $F_{2\text{g}}$ mode of diamond. (b) Evolution of the $F_{2\text{g}}$ mode following THz-SFE and ISRS by a terahertz and visible light pulse, respectively. The envelopes of the excitation pulses are shown schematically. (c) Dependence of the normalized phonon amplitude, $Q/Q_0$, on the duration of the terahertz and visible light pulse, $\tau_\text{THz-SFE}$ and $\tau_\text{ISRS}$. $Q_0$ is the maximum phonon amplitude at $\tau_\text{THz-SFE}=\units{0.2}{ps}$ and $\tau_\text{ISRS}=\units{10}{fs}$. We show two cases, one in which $E_0$ is kept constant while changing $\tau$ (circles), and one in which the total pulse energy is kept constant while changing $\tau$ and $E_0$ accordingly (triangles).
}
\end{figure*}

We begin by reproducing the experiments of references \cite{maehrlein:2017} and \cite{ishioka:2006}, in which the \units{40}{THz} $F_\text{2g}$ Raman-active phonon of diamond was excited via THz-SFE and ISRS, respectively. Both mechanisms can be described by equation~\ref{eq:eom_photonic_R}, for which we use the experimental excitation pulses with a center frequency of $\omega_0/2\pi=\units{20}{THz}$ and pulse duration of $\tau=\units{0.2}{ps}$ for the terahertz pulse for THz-SFE, and $\omega_0/2\pi=\units{375}{THz}~(\units{395}{nm})$ and $\tau=\units{10}{fs}$ for the visible light pulse for ISRS. The electric field is oriented along the Raman-active [110] direction with a peak electric field of $E_0=\units{8}{MV/cm}$ in both cases. The calculated parameters for the equation of motion are given in table~\ref{tab:dft_results}, and the eigenvector of the $F_\text{2g}$ mode is illustrated in figure~\ref{fig:diamond}(a). We show the response of the $F_\text{2g}$ mode to each optical excitation as described by equation~\ref{eq:eom_photonic_R} in figure \ref{fig:diamond}(b).

The response for THz-SFE shows a gradual increase of the phonon amplitude with the onset of the terahertz pulse, which illustrates that the mechanism is nonimpulsive. The maximum phonon amplitude reaches $Q=0.28\times10^{-2}\,\sqrt{\mu}\ang$, and the phase of the oscillation is sensitive to the carrier-envelope phase of the terahertz pulse, $\varphi_\text{CEP}$ \cite{SUPP-SFE:2017,maehrlein:2017}. The response for ISRS shows an abrupt onset of the phonon amplitude at $t=0$, which is characteristic for the impulsive nature of the mechanism. The maximum phonon amplitude reaches $Q=0.04\times10^{-2}\,\sqrt{\mu}\ang$, which agrees well with the results of recent time-dependent density functional theory studies for ISRS \cite{Shinohara2010,Shinohara2:2010}. Here the phase of the oscillation is independent of $\varphi_\text{CEP}$ \cite{SUPP-SFE:2017}.

\begin{table}[b]
\centering
\bgroup
\def\arraystretch{1.3}
\caption{
Calculated phonon frequencies of the infrared- and Raman-active modes, $\Omega_\text{IR}$ and $\Omega_\text{R}$, mode effective charge of the infrared-active mode, $Z_\text{IR}$, Raman tensor at the respective visible and terahertz frequencies of the laser pulses, $R(\omega)$, and quadratic-linear coupling coefficient, $c$.
}
\begin{tabular}{c | c c c}
\hline\hline
Quantity & ~Diamond~ & ~ErFeO$_3$~ & ~BiFeO$_3$~ \\
\hline
$\Omega_\text{IR}$ $|$ $\Omega_\text{R}$ (THz) & -- $|$ 39.2 & 16.5 $|$ 3.2 & 7.4 $|$ 15.3\\
$Z_\text{IR}$ ($e/\sqrt{\mu}$) &  & 0.67 & 0.82 \\
$R$(VIS) ($\ang^2/\sqrt{\mu}$) & 70 &  &  \\
$R$(THz) ($\ang^2/\sqrt{\mu}$) & 50 & -9 & -41 \\
$c$ ($\text{meV}/(\ang\sqrt{\mu})^3$) &  & 7.8 & 8.0\\
\hline\hline
\end{tabular}
\label{tab:dft_results}
\egroup
\end{table}

Despite the smaller value of the Raman tensor $R^{\mathrm{el}}(\text{THz})$ compared to $R^{\mathrm{el}}(\text{VIS})$, see table~\ref{tab:dft_results}, THz-SFE is roughly one order of magnitude stronger than ISRS for a similar electric field strength, $E_0$. This is because the \units{40}{THz} frequency component of the driving force, $E^2(t)$, resulting from the sum frequency of the \units{20}{THz} pulse is roughly a factor of ten higher than that resulting from the difference frequency of the \units{395}{nm} pulse. This order-of-magnitude difference in excitation strength persists for pulses throughout the visible spectrum \cite{SUPP-SFE:2017}.

To take into account the total pulse energy, we show the dependence of the coherent phonon amplitude on the duration of the pump pulse in figure~\ref{fig:diamond}(c) for two distinct cases: (i) constant peak field, $E_0$, and (ii) constant pulse energy. In ISRS there is an optimal value of pulse duration for constant $E_0$ that corresponds to a bandwidth of the pulse, for which the difference-frequency components at \units{40}{THz} are maximal. When the energy of the pulse is fixed, a shorter pulse will trade off for a higher $E_0$ and therefore increase the effect until the pulse gets too short and approaches the single cycle regime. For THz-SFE the situation is different: Due to its nonimpulsive nature, a longer duration of the pulse will continuously increase the coherent phonon amplitude, when $E_0$ is kept constant. In this case, the amplitude will build up until damping, $\kappa$, and excitation force are balanced. In contrast, keeping the total pulse energy constant, a longer pulse will trade off for a lower $E_0$ and the effect decreases.

\subsection{THz-SFE versus ionic Raman Scattering in ErFeO$_3$}

\begin{figure}[t]
\includegraphics[scale=0.225]{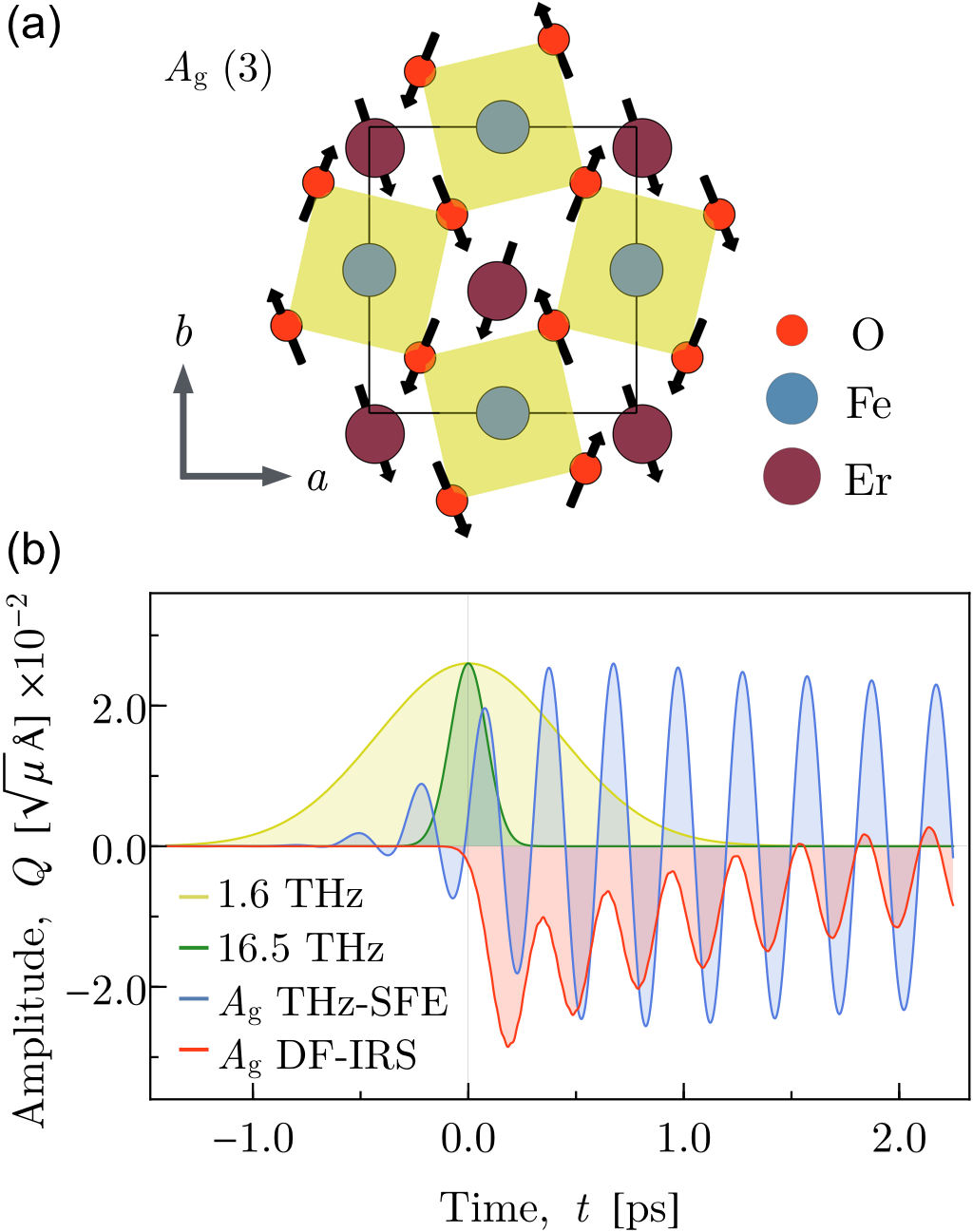}
\caption{
\label{fig:ErFeO3}
(a) Eigenvector of the \units{3.2}{THz} $A_\text{g}$ mode in the $ab$-plane of orthorhombic ErFeO$_3$. The iron ions do not move in this mode. (b) Evolution of the $A_\text{g}$ mode following THz-SFE and DF-IRS by a \units{1.6}{THz} and \units{16.5}{THz} pulse, respectively. The envelopes of the excitation pulses are shown schematically.
}
\end{figure}

We will now compare THz-SFE to the conventional, difference-frequency type of ionic Raman scattering (DF-IRS) at the example of ErFeO$_3$. For orthorhombic ErFeO$_3$ with space group $Pnma$, experimental and theoretical studies are available that show the coupling of the Raman-active \units{3.2}{THz} $A_\text{g}$ mode, see figure~\ref{fig:ErFeO3}(a), with the infrared-active \units{16.5}{THz} $B_\text{3u}$ mode fulfilling the condition $\Omega_\text{IR}>\Omega_\text{R}$ \cite{nova:2017,juraschek:2017}. Therefore, we model two different terahertz pulses, one with $\omega_0=\Omega_\text{R}/2$ for THz-SFE according to equation~\ref{eq:eom_photonic_R}, and the other to initially excite the $B_\text{3u}$ mode for ionic Raman scattering according to equations~\ref{eq:eom_ionic_IR} and \ref{eq:eom_ionic_R}. The electric field for THz-SFE is oriented along the Raman-active $c$ direction with a center frequency of $\omega_0/2\pi=\units{1.6}{THz}$ and pulse duration of $\tau=\units{1}{ps}$; the electric field for DF-IRS is oriented along the infrared-active $a$ direction with $\omega_0/2\pi=\units{16.5}{THz}$ and $\tau=\units{0.2}{ps}$. We assume a peak electric field of $E_0=\units{8}{MV/cm}$ in both cases. The calculated parameters for the equations of motion are given in table~\ref{tab:dft_results}, and the eigenvector of the $A_\text{g}$ mode is illustrated in figure~\ref{fig:ErFeO3}(a). We show the response of the $A_\text{g}$ mode to each optical excitation as described by equations \ref{eq:eom_photonic_R} -- \ref{eq:eom_ionic_R} in figure~\ref{fig:ErFeO3}(b).

As in the case of diamond, the response for THz-SFE shows a continuous increase of the phonon amplitude with the onset of the pulse, reaching a maximum of $Q=2.6\times10^{-2}\,\sqrt{\mu}\ang$. The response for DF-IRS shows an impulsive onset of the phonon amplitude at $t=0$ that is not sensitive to the carrier-envelope phase, $\varphi_\text{CEP}$ \cite{SUPP-SFE:2017}, as well as the typical displacive feature of nonlinear phononics \cite{forst:2011,subedi:2014,juraschek:2017}. In this case, the maximum phonon amplitude reaches $Q=0.9\times10^{-2}\,\sqrt{\mu}\ang$.

The amplitude of the $A_\text{g}$ mode induced by THz-SFE is higher than the amplitude induced by DF-IRS by a factor of three. The excitation mechanisms are fundamentally different however, and the decisive factors are the values of the Raman tensor, $R$, arising from a change in polarizability, and the coupling coefficient, $c$, arising from an anharmonic phonon potential. The comparison between the excitation strengths therefore has to be done for each material and phonon mode. Note that DF-IRS is mainly used because of its unipolar displacive feature.

\subsection{\label{sec:bifeo3}Photonic and ionic sum-frequency excitation in BiFeO$_3$}

\begin{figure*}[t]
\includegraphics[scale=0.225]{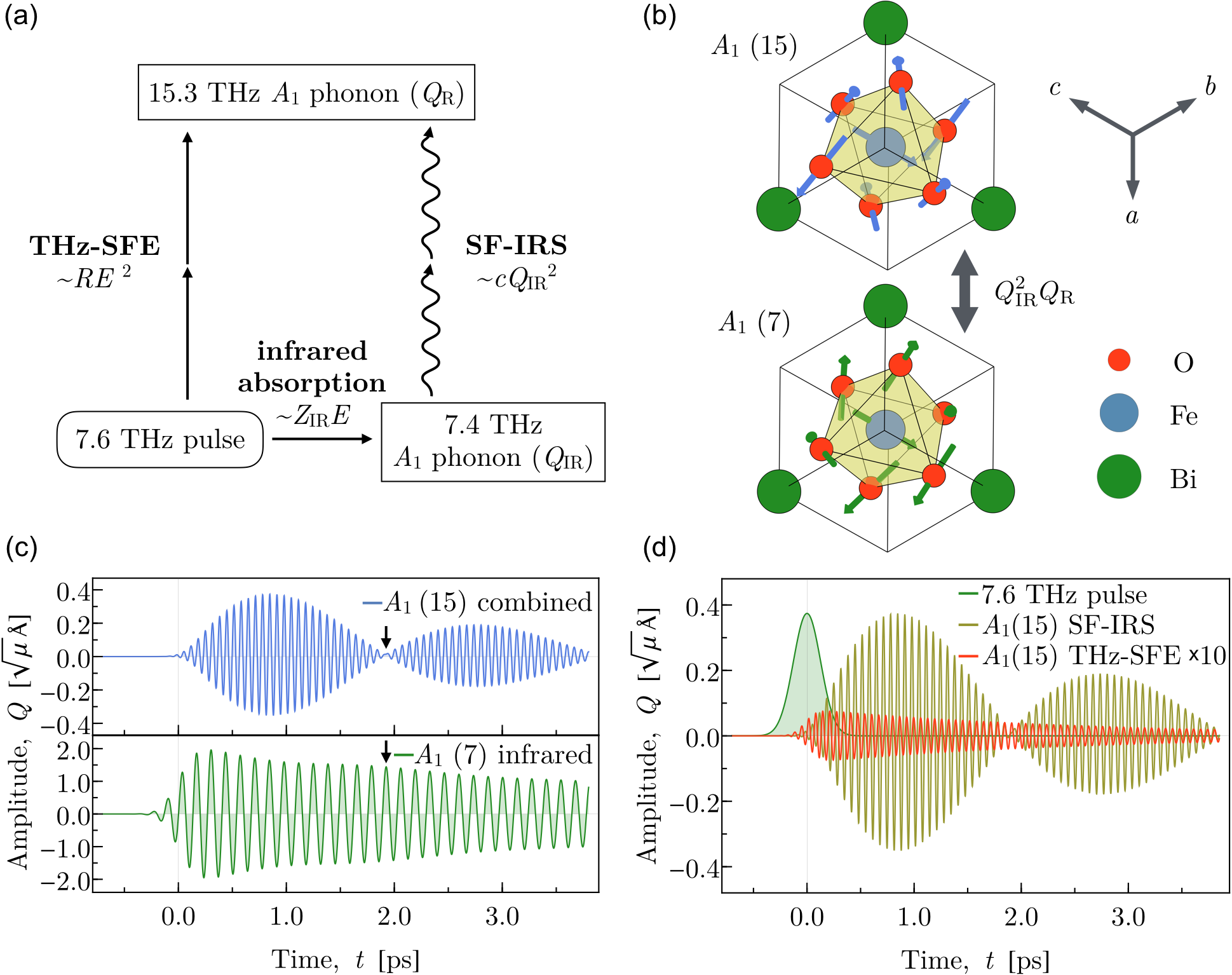}
\caption{
\label{fig:BiFeO3}
(a) Schematic of the excitation by the terahertz pulse. (b) Eigenvectors of the coupled \units{15.3}{THz} (blue) and \units{7.4}{THz} (green) $A_1$ modes from a view along the [111] direction of rhombohedral BiFeO$_3$. The motion of bismuth ions in these modes is negligible. (b) Full evolution of the $A_\text{1}$ modes following the excitation by a single \units{7.6}{THz} pulse. The $A_1 (7)$ mode is excited via infrared absorption. The $A_1 (15)$ mode is excited \textit{simultaneously} via THz-SFE and SF-IRS. Black arrows point to the beat nodes of the $A_1 (15)$ mode that correspond to beat maxima of the $A_1 (7)$ mode. (c) Evolution of the $A_1 (15)$ mode as in (b), but separating the effects of the two mechanisms. The marginal THz-SFE contribution is magnified by a factor of ten for better visibility. The envelope of the excitation pulse is shown schematically.
}
\end{figure*}

In the previous sections we compared the recently demonstrated THz-SFE with the commonly used ISRS and DF-IRS. In this final step, we propose the so-far overlooked sum-frequency counterpart of ionic Raman scattering (SF-IRS) as depicted in figure~\ref{fig:mechanisms}(d). We demonstrate this mechanism and compare it to THz-SFE using the example of BiFeO$_3$. In noncentrosymmetric rhombohedral BiFeO$_3$ with space group $R3c$, all fully symmetric $A_1$ modes are both infrared-active and Raman-active along the [111] direction of the crystal. These modes couple quadratic-linearly to each other, and two of them lie at frequencies of \units{15.3}{THz} and \units{7.4}{THz}. Thus by exciting the system with a single pulse with a center frequency of $\omega_0/2\pi=\units{7.6}{THz}$ we expect both THz-SFE and SF-IRS to occur at the same time: The pulse directly excites the $A_1 (7)$ mode via infrared absorption, which then mediates energy to the $A_1 (15)$ mode via SF-IRS. Simultaneously, the pulse excites the $A_1 (15)$ mode via THz-SFE (but not via infrared absorption as the $A_1 (15)$ phonon lies well outside the \units{2.9}{THz} bandwidth of the pulse). To make this process clearer, we show a schematic of the excitations in figure~\ref{fig:BiFeO3}(a). Note that one could also drive the \units{15.3}{THz} mode directly via infrared absorption. For consistency to the previous sections, we label the \units{7.4}{THz} mode as ``IR'' and the \units{15.3}{THz} mode as ``R'', and both criteria, $\omega_0 \approx \Omega_\text{R}/2$ and $\Omega_\text{IR} \approx \Omega_\text{R}/2$ are fulfilled. We model the terahertz pulse with a center frequency of $\omega_0/2\pi=\units{7.6}{THz}$ and a duration of $\tau=\units{0.3}{ps}$. The electric field is oriented along the Raman- and infrared-active [111] direction with a peak of $E_0=\units{8}{MV/cm}$. The calculated parameters for the equations of motion are given in table~\ref{tab:dft_results}, and the eigenvectors of the $A_1$ modes are illustrated in figure~\ref{fig:BiFeO3}(b). We show the responses of both $A_1$ modes to the optical excitation as described by equations \ref{eq:eom_photonic_R} -- \ref{eq:eom_ionic_R} in figures~\ref{fig:BiFeO3}(c),(d).

The response for the \textit{simultaneous} excitation via THz-SFE and SF-IRS in figure~\ref{fig:BiFeO3}(c) shows a fundamentally different behavior from the other three mechanisms: A beat signal arises and the phonon amplitude reaches by far the highest value of all three examples, $Q=0.37\,\sqrt{\mu}\ang$. We separate the effects of THz-SFE and SF-IRS in figure~\ref{fig:BiFeO3}(d). The response for SF-IRS entirely captures the new feature, while the THz-SFE contribution leads only to a negligible increase of the phonon amplitude and phase shift. The beat signal is caused by a mutual exchange of energy between the infrared-active and Raman-active phonon, and the beat frequency is determined by the strength of the anharmonic phonon coupling, $c$, and through $Q_\text{R}$ by the strength of the terahertz pulse, see equation~\ref{eq:eom_ionic_IR}. A node of the beat signal of the $A_1 (15)$ mode corresponds to a maximum of the beat signal of the $A_1 (7)$ mode, see black arrows in figure~\ref{fig:BiFeO3}(c) - the maximum is swallowed by the damping, however. Naturally, for sum-frequency excitation, the phase of the response is sensitive to the carrier-envelope phase of the terahertz pulse, $\varphi_\text{CEP}$ \cite{SUPP-SFE:2017}. Higher-order anharmonicities in the potential $V(Q_\text{R},Q_\text{IR})$ affect the amplitude, beat-frequency and even introduce new beats, however to a much smaller degree than the quadratic-linear coupling, $Q_\text{IR}^2Q_\text{R}$ \cite{SUPP-SFE:2017}.

\vspace{-0.1cm}


\section{Discussion}

We completed the map of photonic and ionic Raman scattering for the excitation of Raman-active phonons in insulators with the missing sum-frequency part of ionic Raman scattering, see figure~\ref{fig:mechanisms}(d). The difference-frequency mechanisms are impulsive in nature and not sensitive to the carrier-envelope phase of the driving field, $\varphi_\text{CEP}$, whereas the sum-frequency mechanisms are nonimpulsive and therefore sensitive to $\varphi_\text{CEP}$. A summary of the properties is shown in table~\ref{tab:mechanismsproperties}. 

Among the investigated phonon excitations, the up-conversion of frequency components of the driving force is more efficient than the down-conversion. An increase of the total pulse energy will only enhance difference-frequency excitation if it is due to a higher peak electric field, $E_0$, but not due to a longer pulse duration, $\tau$. In contrast, an increase of either $E_0$ or $\tau$ leads to a stronger sum-frequency excitation. This property is particularly relevant for narrowband excitation pulses, for example generated by accelerator-based mid-infrared and terahertz sources \cite{Dhillon2017, Green2016, Gruene2008}. The resulting frequency components are weighted by the Raman tensor and anharmonic phonon coupling, which both depend on the material properties.

For homonuclear materials that do not possess infrared-active phonons, such as diamond, only photonic difference- and sum-frequency excitation is possible. Here the more efficient conversion of sum-frequency components also leads to a higher selectivity for THz-SFE compared to ISRS. Generally, the selectivity depends on the symmetries and frequencies of the phonon modes in the material. In the photonic Raman mechanisms the electric field has to be oriented along the Raman-active direction of the target $Q_\text{R}$ mode, while in the ionic Raman mechanisms it has to be oriented along the infrared-active direction of the coupling $Q_\text{IR}$ mode. Consequently, the selectivity depends on whether ``unwanted'' phonon modes lie within the bandwidth and polarization direction of the driving force $E(t)$ (infrared-active) or $E^2(t)$ (Raman-active) in addition to our target $Q_\text{IR}$ and $Q_\text{R}$ modes. For lattice driven phenomena in the electronic ground state, all three terahertz excitation mechanisms, THz-SFE, DF- and SF-IRS, are favorable over commonly used visible-light or near-infrared ISRS in order to avoid parasitic electronic excitations. The sum-frequency processes provide an additional route to excite optical phonons in the range of 5-15\,THz, for which powerful sources are only now becoming feasible \cite{Dhillon2017,Liu2017}.

With the increased availability of strong terahertz and mid-infrared sources, we anticipate that the presented map of photonic and ionic Raman mechanisms will serve as guide for the selective excitation of crystal lattice vibrations in future. Specifically, we expect that strong excitation of Raman-active phonons will complement the effects arising from infrared-active phonons in the context of spin-phonon and electron-phonon coupled phenomena.


\begin{acknowledgments}
We thank M. Fechner, X.-Y. Zhu, T. Kampfrath, Q. Meier, and N. Spaldin for useful discussions and support. This work was funded by the ETH Z\"{u}rich and ERC-H2020 Grant TERAMAG/No. 681917 (to T. Kampfrath). The work at Columbia University was supported by Office of Naval Research grant \# 01-N00014-18-1-2080 (to X.-Y. Zhu). Calculations were performed at the Swiss National Supercomputing Centre (CSCS) supported by the project IDs s624 and p504. 
\end{acknowledgments}



\begin{thebibliography}{49}%
\makeatletter
\providecommand \@ifxundefined [1]{%
 \@ifx{#1\undefined}
}%
\providecommand \@ifnum [1]{%
 \ifnum #1\expandafter \@firstoftwo
 \else \expandafter \@secondoftwo
 \fi
}%
\providecommand \@ifx [1]{%
 \ifx #1\expandafter \@firstoftwo
 \else \expandafter \@secondoftwo
 \fi
}%
\providecommand \natexlab [1]{#1}%
\providecommand \enquote  [1]{``#1''}%
\providecommand \bibnamefont  [1]{#1}%
\providecommand \bibfnamefont [1]{#1}%
\providecommand \citenamefont [1]{#1}%
\providecommand \href@noop [0]{\@secondoftwo}%
\providecommand \href [0]{\begingroup \@sanitize@url \@href}%
\providecommand \@href[1]{\@@startlink{#1}\@@href}%
\providecommand \@@href[1]{\endgroup#1\@@endlink}%
\providecommand \@sanitize@url [0]{\catcode `\\12\catcode `\$12\catcode
  `\&12\catcode `\#12\catcode `\^12\catcode `\_12\catcode `\%12\relax}%
\providecommand \@@startlink[1]{}%
\providecommand \@@endlink[0]{}%
\providecommand \url  [0]{\begingroup\@sanitize@url \@url }%
\providecommand \@url [1]{\endgroup\@href {#1}{\urlprefix }}%
\providecommand \urlprefix  [0]{URL }%
\providecommand \Eprint [0]{\href }%
\providecommand \doibase [0]{http://dx.doi.org/}%
\providecommand \selectlanguage [0]{\@gobble}%
\providecommand \bibinfo  [0]{\@secondoftwo}%
\providecommand \bibfield  [0]{\@secondoftwo}%
\providecommand \translation [1]{[#1]}%
\providecommand \BibitemOpen [0]{}%
\providecommand \bibitemStop [0]{}%
\providecommand \bibitemNoStop [0]{.\EOS\space}%
\providecommand \EOS [0]{\spacefactor3000\relax}%
\providecommand \BibitemShut  [1]{\csname bibitem#1\endcsname}%
\let\auto@bib@innerbib\@empty
\bibitem [{\citenamefont {Juraschek}\ and\ \citenamefont
  {Spaldin}(2017)}]{juraschek3:2017}%
  \BibitemOpen
  \bibfield  {author} {\bibinfo {author} {\bibfnamefont {D.~M.}\ \bibnamefont
  {Juraschek}}\ and\ \bibinfo {author} {\bibfnamefont {N.~A.}\ \bibnamefont
  {Spaldin}},\ }\bibfield  {title} {\enquote {\bibinfo {title} {{Sounding out
  optical phonons}},}\ }\href@noop {} {\bibfield  {journal} {\bibinfo
  {journal} {Science}\ }\textbf {\bibinfo {volume} {357}},\ \bibinfo {pages}
  {9--10} (\bibinfo {year} {2017})}\BibitemShut {NoStop}%
\bibitem [{\citenamefont {Kampfrath}\ \emph {et~al.}(2013)\citenamefont
  {Kampfrath}, \citenamefont {Tanaka},\ and\ \citenamefont
  {Nelson}}]{Kampfrath2013a}%
  \BibitemOpen
  \bibfield  {author} {\bibinfo {author} {\bibfnamefont {T.}~\bibnamefont
  {Kampfrath}}, \bibinfo {author} {\bibfnamefont {K.}~\bibnamefont {Tanaka}}, \
  and\ \bibinfo {author} {\bibfnamefont {K.~A.}\ \bibnamefont {Nelson}},\
  }\bibfield  {title} {\enquote {\bibinfo {title} {Resonant and nonresonant
  control over matter and light by intense terahertz transients},}\ }\href
  {http://dx.doi.org/10.1038/nphoton.2013.184} {\bibfield  {journal} {\bibinfo
  {journal} {Nat. Photonics}\ }\textbf {\bibinfo {volume} {7}},\ \bibinfo
  {pages} {680--690} (\bibinfo {year} {2013})}\BibitemShut {NoStop}%
\bibitem [{\citenamefont {Dhillon}\ \emph {et~al.}(2017)\citenamefont
  {Dhillon}, \citenamefont {Vitiello}, \citenamefont {Linfield}, \citenamefont
  {Davies}, \citenamefont {Hoffmann}, \citenamefont {Booske}, \citenamefont
  {Paoloni}, \citenamefont {Gensch}, \citenamefont {Weightman}, \citenamefont
  {Williams}, \citenamefont {Castro-Camus}, \citenamefont {Cumming},
  \citenamefont {Simoens}, \citenamefont {Escorcia-Carranza}, \citenamefont
  {Grant}, \citenamefont {Lucyszyn}, \citenamefont {Kuwata-Gonokami},
  \citenamefont {Konishi}, \citenamefont {Koch}, \citenamefont {Schmuttenmaer},
  \citenamefont {Cocker}, \citenamefont {Huber}, \citenamefont {Markelz},
  \citenamefont {Taylor}, \citenamefont {Wallace}, \citenamefont {{Axel
  Zeitler}}, \citenamefont {Sibik}, \citenamefont {Korter}, \citenamefont
  {Ellison}, \citenamefont {Rea}, \citenamefont {Goldsmith}, \citenamefont
  {Cooper}, \citenamefont {Appleby}, \citenamefont {Pardo}, \citenamefont
  {Huggard}, \citenamefont {Krozer}, \citenamefont {Shams}, \citenamefont
  {Fice}, \citenamefont {Renaud}, \citenamefont {Seeds}, \citenamefont
  {St{\"{o}}hr}, \citenamefont {Naftaly}, \citenamefont {Ridler}, \citenamefont
  {Clarke}, \citenamefont {Cunningham},\ and\ \citenamefont
  {Johnston}}]{Dhillon2017}%
  \BibitemOpen
  \bibfield  {author} {\bibinfo {author} {\bibfnamefont {S.~S.}\ \bibnamefont
  {Dhillon}}, \bibinfo {author} {\bibfnamefont {M.~S.}\ \bibnamefont
  {Vitiello}}, \bibinfo {author} {\bibfnamefont {E.~H.}\ \bibnamefont
  {Linfield}}, \bibinfo {author} {\bibfnamefont {A.~G.}\ \bibnamefont
  {Davies}}, \bibinfo {author} {\bibfnamefont {Matthias~C.}\ \bibnamefont
  {Hoffmann}}, \bibinfo {author} {\bibfnamefont {J.}~\bibnamefont {Booske}},
  \bibinfo {author} {\bibfnamefont {C.}~\bibnamefont {Paoloni}}, \bibinfo
  {author} {\bibfnamefont {M.}~\bibnamefont {Gensch}}, \bibinfo {author}
  {\bibfnamefont {P.}~\bibnamefont {Weightman}}, \bibinfo {author}
  {\bibfnamefont {G.~P.}\ \bibnamefont {Williams}}, \bibinfo {author}
  {\bibfnamefont {E.}~\bibnamefont {Castro-Camus}}, \bibinfo {author}
  {\bibfnamefont {D.~R.S.}\ \bibnamefont {Cumming}}, \bibinfo {author}
  {\bibfnamefont {F.}~\bibnamefont {Simoens}}, \bibinfo {author} {\bibfnamefont
  {I.}~\bibnamefont {Escorcia-Carranza}}, \bibinfo {author} {\bibfnamefont
  {J.}~\bibnamefont {Grant}}, \bibinfo {author} {\bibfnamefont
  {S.}~\bibnamefont {Lucyszyn}}, \bibinfo {author} {\bibfnamefont
  {M.}~\bibnamefont {Kuwata-Gonokami}}, \bibinfo {author} {\bibfnamefont
  {K.}~\bibnamefont {Konishi}}, \bibinfo {author} {\bibfnamefont
  {M.}~\bibnamefont {Koch}}, \bibinfo {author} {\bibfnamefont {C.~A.}\
  \bibnamefont {Schmuttenmaer}}, \bibinfo {author} {\bibfnamefont {T.~L.}\
  \bibnamefont {Cocker}}, \bibinfo {author} {\bibfnamefont {R.}~\bibnamefont
  {Huber}}, \bibinfo {author} {\bibfnamefont {A.~G.}\ \bibnamefont {Markelz}},
  \bibinfo {author} {\bibfnamefont {Z.~D.}\ \bibnamefont {Taylor}}, \bibinfo
  {author} {\bibfnamefont {V.~P.}\ \bibnamefont {Wallace}}, \bibinfo {author}
  {\bibfnamefont {J.}~\bibnamefont {{Axel Zeitler}}}, \bibinfo {author}
  {\bibfnamefont {J.}~\bibnamefont {Sibik}}, \bibinfo {author} {\bibfnamefont
  {T.~M.}\ \bibnamefont {Korter}}, \bibinfo {author} {\bibfnamefont
  {B.}~\bibnamefont {Ellison}}, \bibinfo {author} {\bibfnamefont
  {S.}~\bibnamefont {Rea}}, \bibinfo {author} {\bibfnamefont {P.}~\bibnamefont
  {Goldsmith}}, \bibinfo {author} {\bibfnamefont {K.~B.}\ \bibnamefont
  {Cooper}}, \bibinfo {author} {\bibfnamefont {R.}~\bibnamefont {Appleby}},
  \bibinfo {author} {\bibfnamefont {D.}~\bibnamefont {Pardo}}, \bibinfo
  {author} {\bibfnamefont {P.~G.}\ \bibnamefont {Huggard}}, \bibinfo {author}
  {\bibfnamefont {V.}~\bibnamefont {Krozer}}, \bibinfo {author} {\bibfnamefont
  {H.}~\bibnamefont {Shams}}, \bibinfo {author} {\bibfnamefont
  {M.}~\bibnamefont {Fice}}, \bibinfo {author} {\bibfnamefont {C.}~\bibnamefont
  {Renaud}}, \bibinfo {author} {\bibfnamefont {A.}~\bibnamefont {Seeds}},
  \bibinfo {author} {\bibfnamefont {A.}~\bibnamefont {St{\"{o}}hr}}, \bibinfo
  {author} {\bibfnamefont {M.}~\bibnamefont {Naftaly}}, \bibinfo {author}
  {\bibfnamefont {N.}~\bibnamefont {Ridler}}, \bibinfo {author} {\bibfnamefont
  {R.}~\bibnamefont {Clarke}}, \bibinfo {author} {\bibfnamefont {J.~E.}\
  \bibnamefont {Cunningham}}, \ and\ \bibinfo {author} {\bibfnamefont {M.~B.}\
  \bibnamefont {Johnston}},\ }\bibfield  {title} {\enquote {\bibinfo {title}
  {{The 2017 terahertz science and technology roadmap}},}\ }\href@noop {}
  {\bibfield  {journal} {\bibinfo  {journal} {J.~Phys. D: Appl. Phys.}\
  }\textbf {\bibinfo {volume} {50}} (\bibinfo {year} {2017})}\BibitemShut
  {NoStop}%
\bibitem [{\citenamefont {Sell}\ \emph {et~al.}(2008)\citenamefont {Sell},
  \citenamefont {Leitenstorfer},\ and\ \citenamefont {Huber}}]{Sell2008}%
  \BibitemOpen
  \bibfield  {author} {\bibinfo {author} {\bibfnamefont {A.}~\bibnamefont
  {Sell}}, \bibinfo {author} {\bibfnamefont {A.}~\bibnamefont {Leitenstorfer}},
  \ and\ \bibinfo {author} {\bibfnamefont {R.}~\bibnamefont {Huber}},\
  }\bibfield  {title} {\enquote {\bibinfo {title} {Phase-locked generation and
  field-resolved detection of widely tunable terahertz pulses with amplitudes
  exceeding 100 {MV}/cm},}\ }\href
  {http://ol.osa.org/abstract.cfm?URI=ol-33-23-2767} {\bibfield  {journal}
  {\bibinfo  {journal} {Opt. Lett.}\ }\textbf {\bibinfo {volume} {33}},\
  \bibinfo {pages} {2767--2769} (\bibinfo {year} {2008})}\BibitemShut {NoStop}%
\bibitem [{\citenamefont {Rini}\ \emph {et~al.}(2007)\citenamefont {Rini},
  \citenamefont {Tobey}, \citenamefont {Dean}, \citenamefont {Itatani},
  \citenamefont {Tomioka}, \citenamefont {Tokura}, \citenamefont {Schoenlein},\
  and\ \citenamefont {Cavalleri}}]{Rini2007}%
  \BibitemOpen
  \bibfield  {author} {\bibinfo {author} {\bibfnamefont {M.}~\bibnamefont
  {Rini}}, \bibinfo {author} {\bibfnamefont {R.}~\bibnamefont {Tobey}},
  \bibinfo {author} {\bibfnamefont {N.}~\bibnamefont {Dean}}, \bibinfo {author}
  {\bibfnamefont {J.}~\bibnamefont {Itatani}}, \bibinfo {author} {\bibfnamefont
  {Y.}~\bibnamefont {Tomioka}}, \bibinfo {author} {\bibfnamefont
  {Y.}~\bibnamefont {Tokura}}, \bibinfo {author} {\bibfnamefont {R.~W.}\
  \bibnamefont {Schoenlein}}, \ and\ \bibinfo {author} {\bibfnamefont
  {A.}~\bibnamefont {Cavalleri}},\ }\bibfield  {title} {\enquote {\bibinfo
  {title} {{Control of the electronic phase of a manganite by mode-selective
  vibrational excitation.}}}\ }\href@noop {} {\bibfield  {journal} {\bibinfo
  {journal} {Nature}\ }\textbf {\bibinfo {volume} {449}},\ \bibinfo {pages}
  {72--74} (\bibinfo {year} {2007})}\BibitemShut {NoStop}%
\bibitem [{\citenamefont {Fausti}\ \emph {et~al.}(2011)\citenamefont {Fausti},
  \citenamefont {Tobey}, \citenamefont {Dean}, \citenamefont {Kaiser},
  \citenamefont {Dienst}, \citenamefont {Hoffmann}, \citenamefont {Pyon},
  \citenamefont {Takayama}, \citenamefont {Takagi},\ and\ \citenamefont
  {Cavalleri}}]{fausti:2011}%
  \BibitemOpen
  \bibfield  {author} {\bibinfo {author} {\bibfnamefont {D.}~\bibnamefont
  {Fausti}}, \bibinfo {author} {\bibfnamefont {R.~I.}\ \bibnamefont {Tobey}},
  \bibinfo {author} {\bibfnamefont {N.}~\bibnamefont {Dean}}, \bibinfo {author}
  {\bibfnamefont {S.}~\bibnamefont {Kaiser}}, \bibinfo {author} {\bibfnamefont
  {A.}~\bibnamefont {Dienst}}, \bibinfo {author} {\bibfnamefont {M.~C.}\
  \bibnamefont {Hoffmann}}, \bibinfo {author} {\bibfnamefont {S.}~\bibnamefont
  {Pyon}}, \bibinfo {author} {\bibfnamefont {T.}~\bibnamefont {Takayama}},
  \bibinfo {author} {\bibfnamefont {H.}~\bibnamefont {Takagi}}, \ and\ \bibinfo
  {author} {\bibfnamefont {A.}~\bibnamefont {Cavalleri}},\ }\bibfield  {title}
  {\enquote {\bibinfo {title} {Light-induced superconductivity in a
  stripe-ordered cuprate},}\ }\href@noop {} {\bibfield  {journal} {\bibinfo
  {journal} {Science}\ }\textbf {\bibinfo {volume} {331}},\ \bibinfo {pages}
  {189} (\bibinfo {year} {2011})}\BibitemShut {NoStop}%
\bibitem [{\citenamefont {Hu}\ \emph {et~al.}(2014)\citenamefont {Hu},
  \citenamefont {Kaiser}, \citenamefont {Nicoletti}, \citenamefont {Hunt},
  \citenamefont {Gierz}, \citenamefont {Hoffmann}, \citenamefont {{Le Tacon}},
  \citenamefont {Loew}, \citenamefont {Keimer},\ and\ \citenamefont
  {Cavalleri}}]{hu:2014}%
  \BibitemOpen
  \bibfield  {author} {\bibinfo {author} {\bibfnamefont {W.}~\bibnamefont
  {Hu}}, \bibinfo {author} {\bibfnamefont {S.}~\bibnamefont {Kaiser}}, \bibinfo
  {author} {\bibfnamefont {D.}~\bibnamefont {Nicoletti}}, \bibinfo {author}
  {\bibfnamefont {C.~R.}\ \bibnamefont {Hunt}}, \bibinfo {author}
  {\bibfnamefont {I.}~\bibnamefont {Gierz}}, \bibinfo {author} {\bibfnamefont
  {M.~C.}\ \bibnamefont {Hoffmann}}, \bibinfo {author} {\bibfnamefont
  {M.}~\bibnamefont {{Le Tacon}}}, \bibinfo {author} {\bibfnamefont
  {T.}~\bibnamefont {Loew}}, \bibinfo {author} {\bibfnamefont {B.}~\bibnamefont
  {Keimer}}, \ and\ \bibinfo {author} {\bibfnamefont {A.}~\bibnamefont
  {Cavalleri}},\ }\bibfield  {title} {\enquote {\bibinfo {title} {Optically
  enhanced coherent transport in {YBa}$_2${Cu}$_3${O}$_{6.5}$ by ultrafast
  redistribution of interlayer coupling},}\ }\href@noop {} {\bibfield
  {journal} {\bibinfo  {journal} {Nat. Mater.}\ }\textbf {\bibinfo {volume}
  {13}},\ \bibinfo {pages} {705} (\bibinfo {year} {2014})}\BibitemShut
  {NoStop}%
\bibitem [{\citenamefont {Mitrano}\ \emph {et~al.}(2016)\citenamefont
  {Mitrano}, \citenamefont {Cantaluppi}, \citenamefont {Nicoletti},
  \citenamefont {Kaiser}, \citenamefont {Perucchi}, \citenamefont {Lupi},
  \citenamefont {{Di Pietro}}, \citenamefont {Pontiroli}, \citenamefont
  {Ricc{\`{o}}}, \citenamefont {Clark}, \citenamefont {Jaksch},\ and\
  \citenamefont {Cavalleri}}]{Mitrano2016}%
  \BibitemOpen
  \bibfield  {author} {\bibinfo {author} {\bibfnamefont {M.}~\bibnamefont
  {Mitrano}}, \bibinfo {author} {\bibfnamefont {A.}~\bibnamefont {Cantaluppi}},
  \bibinfo {author} {\bibfnamefont {D.}~\bibnamefont {Nicoletti}}, \bibinfo
  {author} {\bibfnamefont {S.}~\bibnamefont {Kaiser}}, \bibinfo {author}
  {\bibfnamefont {A.}~\bibnamefont {Perucchi}}, \bibinfo {author}
  {\bibfnamefont {S.}~\bibnamefont {Lupi}}, \bibinfo {author} {\bibfnamefont
  {P.}~\bibnamefont {{Di Pietro}}}, \bibinfo {author} {\bibfnamefont
  {D.}~\bibnamefont {Pontiroli}}, \bibinfo {author} {\bibfnamefont
  {M.}~\bibnamefont {Ricc{\`{o}}}}, \bibinfo {author} {\bibfnamefont {S.~R.}\
  \bibnamefont {Clark}}, \bibinfo {author} {\bibfnamefont {D.}~\bibnamefont
  {Jaksch}}, \ and\ \bibinfo {author} {\bibfnamefont {A.}~\bibnamefont
  {Cavalleri}},\ }\bibfield  {title} {\enquote {\bibinfo {title} {{Possible
  light-induced superconductivity in K$_3$C$_{60}$ at high temperature}},}\
  }\href@noop {} {\bibfield  {journal} {\bibinfo  {journal} {Nature}\ }\textbf
  {\bibinfo {volume} {530}},\ \bibinfo {pages} {461--464} (\bibinfo {year}
  {2016})}\BibitemShut {NoStop}%
\bibitem [{\citenamefont {Nova}\ \emph {et~al.}(2017)\citenamefont {Nova},
  \citenamefont {Cartella}, \citenamefont {Cantaluppi}, \citenamefont
  {F{\"{o}}rst}, \citenamefont {Bossini}, \citenamefont {Mikhaylovskiy},
  \citenamefont {Kimel}, \citenamefont {Merlin},\ and\ \citenamefont
  {Cavalleri}}]{nova:2017}%
  \BibitemOpen
  \bibfield  {author} {\bibinfo {author} {\bibfnamefont {T.~F.}\ \bibnamefont
  {Nova}}, \bibinfo {author} {\bibfnamefont {A.}~\bibnamefont {Cartella}},
  \bibinfo {author} {\bibfnamefont {A.}~\bibnamefont {Cantaluppi}}, \bibinfo
  {author} {\bibfnamefont {M.}~\bibnamefont {F{\"{o}}rst}}, \bibinfo {author}
  {\bibfnamefont {D.}~\bibnamefont {Bossini}}, \bibinfo {author} {\bibfnamefont
  {R.~V.}\ \bibnamefont {Mikhaylovskiy}}, \bibinfo {author} {\bibfnamefont
  {A.~V.}\ \bibnamefont {Kimel}}, \bibinfo {author} {\bibfnamefont
  {R.}~\bibnamefont {Merlin}}, \ and\ \bibinfo {author} {\bibfnamefont
  {A.}~\bibnamefont {Cavalleri}},\ }\bibfield  {title} {\enquote {\bibinfo
  {title} {{An effective magnetic field from optically driven phonons}},}\
  }\href@noop {} {\bibfield  {journal} {\bibinfo  {journal} {Nat. Phys.}\
  }\textbf {\bibinfo {volume} {13}},\ \bibinfo {pages} {132--137} (\bibinfo
  {year} {2017})}\BibitemShut {NoStop}%
\bibitem [{\citenamefont {Maehrlein}\ \emph
  {et~al.}(2017{\natexlab{a}})\citenamefont {Maehrlein}, \citenamefont {Radu},
  \citenamefont {Maldonado}, \citenamefont {Paarmann}, \citenamefont {Gensch},
  \citenamefont {Kalashnikova}, \citenamefont {Pisarev}, \citenamefont {Wolf},
  \citenamefont {Oppeneer}, \citenamefont {Barker},\ and\ \citenamefont
  {Kampfrath}}]{Maehrlein2018}%
  \BibitemOpen
  \bibfield  {author} {\bibinfo {author} {\bibfnamefont {S.}~\bibnamefont
  {Maehrlein}}, \bibinfo {author} {\bibfnamefont {I.}~\bibnamefont {Radu}},
  \bibinfo {author} {\bibfnamefont {P.}~\bibnamefont {Maldonado}}, \bibinfo
  {author} {\bibfnamefont {A.}~\bibnamefont {Paarmann}}, \bibinfo {author}
  {\bibfnamefont {M.}~\bibnamefont {Gensch}}, \bibinfo {author} {\bibfnamefont
  {A.~M.}\ \bibnamefont {Kalashnikova}}, \bibinfo {author} {\bibfnamefont
  {R.~V.}\ \bibnamefont {Pisarev}}, \bibinfo {author} {\bibfnamefont
  {M.}~\bibnamefont {Wolf}}, \bibinfo {author} {\bibfnamefont {P.~M.}\
  \bibnamefont {Oppeneer}}, \bibinfo {author} {\bibfnamefont {J.}~\bibnamefont
  {Barker}}, \ and\ \bibinfo {author} {\bibfnamefont {T.}~\bibnamefont
  {Kampfrath}},\ }\bibfield  {title} {\enquote {\bibinfo {title} {Revealing
  spin-phonon interaction in ferrimagnetic insulators by ultrafast lattice
  excitation},}\ }\href@noop {} {\bibfield  {journal} {\bibinfo  {journal}
  {ArXiv e-prints}\ } (\bibinfo {year} {2017}{\natexlab{a}})},\ \Eprint
  {http://arxiv.org/abs/1710.02700} {arXiv:1710.02700} \BibitemShut {NoStop}%
\bibitem [{\citenamefont {Dekorsy}\ \emph {et~al.}(2000)\citenamefont
  {Dekorsy}, \citenamefont {Cho},\ and\ \citenamefont {Kurz}}]{Dekorsy2000}%
  \BibitemOpen
  \bibfield  {author} {\bibinfo {author} {\bibfnamefont {T.}~\bibnamefont
  {Dekorsy}}, \bibinfo {author} {\bibfnamefont {G.~C.}\ \bibnamefont {Cho}}, \
  and\ \bibinfo {author} {\bibfnamefont {H.}~\bibnamefont {Kurz}},\ }\bibfield
  {title} {\enquote {\bibinfo {title} {Coherent phonons in condensed media},}\
  }in\ \href {\doibase 10.1007/BFb0084242} {\emph {\bibinfo {booktitle} {Topics
  in Applied Physics}}},\ Vol.~\bibinfo {volume} {76},\ \bibinfo {editor}
  {edited by\ \bibinfo {editor} {\bibfnamefont {M.}~\bibnamefont {Cardona}}\
  and\ \bibinfo {editor} {\bibfnamefont {G.}~\bibnamefont {G\"untherodt}}}\
  (\bibinfo  {publisher} {Springer Berlin Heidelberg},\ \bibinfo {year}
  {2000})\ pp.\ \bibinfo {pages} {169--209}\BibitemShut {NoStop}%
\bibitem [{\citenamefont {F\"{o}rst}\ and\ \citenamefont
  {Dekorsy}(2008)}]{Forst2008}%
  \BibitemOpen
  \bibfield  {author} {\bibinfo {author} {\bibfnamefont {M.}~\bibnamefont
  {F\"{o}rst}}\ and\ \bibinfo {author} {\bibfnamefont {T.}~\bibnamefont
  {Dekorsy}},\ }\bibfield  {title} {\enquote {\bibinfo {title} {{Coherent
  Phonons in Bulk and Low-Dimensional Semiconductors}},}\ }in\ \href@noop {}
  {\emph {\bibinfo {booktitle} {Coherent Vibrational Dynamics}}}\ (\bibinfo
  {publisher} {Taylor \&{} Francis Group},\ \bibinfo {year} {2008})\ pp.\
  \bibinfo {pages} {129--172}\BibitemShut {NoStop}%
\bibitem [{Note1()}]{Note1}%
  \BibitemOpen
  \bibinfo {note} {Under photoexcitation there are a lot more mechanisms that
  excite Raman-active phonons, such as ``displacive excitation of coherent
  phonons'' or ``transient depletion field screening''.}\BibitemShut {Stop}%
\bibitem [{\citenamefont {{De Silvestri}}\ \emph {et~al.}(1985)\citenamefont
  {{De Silvestri}}, \citenamefont {Fujimoto}, \citenamefont {Ippen},
  \citenamefont {{Gamble Jr.}}, \citenamefont {Williams},\ and\ \citenamefont
  {Nelson}}]{desilvestri:1985}%
  \BibitemOpen
  \bibfield  {author} {\bibinfo {author} {\bibfnamefont {S.}~\bibnamefont {{De
  Silvestri}}}, \bibinfo {author} {\bibfnamefont {J.~G.}\ \bibnamefont
  {Fujimoto}}, \bibinfo {author} {\bibfnamefont {E.~P.}\ \bibnamefont {Ippen}},
  \bibinfo {author} {\bibfnamefont {E.~B.}\ \bibnamefont {{Gamble Jr.}}},
  \bibinfo {author} {\bibfnamefont {L.~R.}\ \bibnamefont {Williams}}, \ and\
  \bibinfo {author} {\bibfnamefont {K.~A.}\ \bibnamefont {Nelson}},\ }\bibfield
   {title} {\enquote {\bibinfo {title} {{Femtosecond Time-Resolved Measurements
  of Optic Phonon Dephasing by Impulsive Stimulated Raman Scattering in
  $\alpha$--Perylene Crystal from 20 to 300 K}},}\ }\href@noop {} {\bibfield
  {journal} {\bibinfo  {journal} {Chem. Phys. Lett.}\ }\textbf {\bibinfo
  {volume} {116}},\ \bibinfo {pages} {146--152} (\bibinfo {year}
  {1985})}\BibitemShut {NoStop}%
\bibitem [{\citenamefont {Merlin}(1997)}]{merlin:1997}%
  \BibitemOpen
  \bibfield  {author} {\bibinfo {author} {\bibfnamefont {R.}~\bibnamefont
  {Merlin}},\ }\bibfield  {title} {\enquote {\bibinfo {title} {{Generating
  coherent THz phonons with light pulses}},}\ }\href@noop {} {\bibfield
  {journal} {\bibinfo  {journal} {Solid State Commun.}\ }\textbf {\bibinfo
  {volume} {102}},\ \bibinfo {pages} {207--220} (\bibinfo {year}
  {1997})}\BibitemShut {NoStop}%
\bibitem [{\citenamefont {Stevens}\ \emph {et~al.}(2002)\citenamefont
  {Stevens}, \citenamefont {Kuhl},\ and\ \citenamefont {Merlin}}]{Stevens2002}%
  \BibitemOpen
  \bibfield  {author} {\bibinfo {author} {\bibfnamefont {T.}~\bibnamefont
  {Stevens}}, \bibinfo {author} {\bibfnamefont {J.}~\bibnamefont {Kuhl}}, \
  and\ \bibinfo {author} {\bibfnamefont {R.}~\bibnamefont {Merlin}},\
  }\bibfield  {title} {\enquote {\bibinfo {title} {{Coherent phonon generation
  and the two stimulated Raman tensors}},}\ }\href@noop {} {\bibfield
  {journal} {\bibinfo  {journal} {Phys. Rev. B}\ }\textbf {\bibinfo {volume}
  {65}},\ \bibinfo {pages} {144304} (\bibinfo {year} {2002})}\BibitemShut
  {NoStop}%
\bibitem [{\citenamefont {Maradurin}\ and\ \citenamefont
  {Wallis}(1970)}]{maradurin:1970}%
  \BibitemOpen
  \bibfield  {author} {\bibinfo {author} {\bibfnamefont {A.~A.}\ \bibnamefont
  {Maradurin}}\ and\ \bibinfo {author} {\bibfnamefont {R.~F.}\ \bibnamefont
  {Wallis}},\ }\bibfield  {title} {\enquote {\bibinfo {title} {{Ionic Raman
  Effect. I. Scattering by Localized Vibration Modes}},}\ }\href@noop {}
  {\bibfield  {journal} {\bibinfo  {journal} {Phys. Rev. B}\ }\textbf {\bibinfo
  {volume} {2}},\ \bibinfo {pages} {4294--4299} (\bibinfo {year}
  {1970})}\BibitemShut {NoStop}%
\bibitem [{\citenamefont {F{\"{o}}rst}\ \emph {et~al.}(2011)\citenamefont
  {F{\"{o}}rst}, \citenamefont {Manzoni}, \citenamefont {Kaiser}, \citenamefont
  {Tomioka}, \citenamefont {Tokura}, \citenamefont {Merlin},\ and\
  \citenamefont {Cavalleri}}]{forst:2011}%
  \BibitemOpen
  \bibfield  {author} {\bibinfo {author} {\bibfnamefont {M.}~\bibnamefont
  {F{\"{o}}rst}}, \bibinfo {author} {\bibfnamefont {C.}~\bibnamefont
  {Manzoni}}, \bibinfo {author} {\bibfnamefont {S.}~\bibnamefont {Kaiser}},
  \bibinfo {author} {\bibfnamefont {Y.}~\bibnamefont {Tomioka}}, \bibinfo
  {author} {\bibfnamefont {Y.}~\bibnamefont {Tokura}}, \bibinfo {author}
  {\bibfnamefont {R.}~\bibnamefont {Merlin}}, \ and\ \bibinfo {author}
  {\bibfnamefont {A.}~\bibnamefont {Cavalleri}},\ }\bibfield  {title} {\enquote
  {\bibinfo {title} {{Nonlinear phononics as an ultrafast route to lattice
  control}},}\ }\href@noop {} {\bibfield  {journal} {\bibinfo  {journal} {Nat.
  Phys.}\ }\textbf {\bibinfo {volume} {7}},\ \bibinfo {pages} {854--856}
  (\bibinfo {year} {2011})}\BibitemShut {NoStop}%
\bibitem [{\citenamefont {Subedi}\ \emph {et~al.}(2014)\citenamefont {Subedi},
  \citenamefont {Cavalleri},\ and\ \citenamefont {Georges}}]{subedi:2014}%
  \BibitemOpen
  \bibfield  {author} {\bibinfo {author} {\bibfnamefont {A.}~\bibnamefont
  {Subedi}}, \bibinfo {author} {\bibfnamefont {A.}~\bibnamefont {Cavalleri}}, \
  and\ \bibinfo {author} {\bibfnamefont {A.}~\bibnamefont {Georges}},\
  }\bibfield  {title} {\enquote {\bibinfo {title} {Theory of nonlinear
  phononics for coherent light control of solids},}\ }\href@noop {} {\bibfield
  {journal} {\bibinfo  {journal} {Phys. Rev. B}\ }\textbf {\bibinfo {volume}
  {89}},\ \bibinfo {pages} {220301} (\bibinfo {year} {2014})}\BibitemShut
  {NoStop}%
\bibitem [{\citenamefont {Nicoletti}\ and\ \citenamefont
  {Cavalleri}(2016)}]{nicoletti:2016}%
  \BibitemOpen
  \bibfield  {author} {\bibinfo {author} {\bibfnamefont {D.}~\bibnamefont
  {Nicoletti}}\ and\ \bibinfo {author} {\bibfnamefont {A.}~\bibnamefont
  {Cavalleri}},\ }\bibfield  {title} {\enquote {\bibinfo {title} {{Nonlinear
  light--matter interaction at terahertz frequencies}},}\ }\href@noop {}
  {\bibfield  {journal} {\bibinfo  {journal} {Adv. Opt. Photonics}\ }\textbf
  {\bibinfo {volume} {8}},\ \bibinfo {pages} {401--464} (\bibinfo {year}
  {2016})}\BibitemShut {NoStop}%
\bibitem [{\citenamefont {Maehrlein}\ \emph
  {et~al.}(2017{\natexlab{b}})\citenamefont {Maehrlein}, \citenamefont
  {Paarmann}, \citenamefont {Wolf},\ and\ \citenamefont
  {Kampfrath}}]{maehrlein:2017}%
  \BibitemOpen
  \bibfield  {author} {\bibinfo {author} {\bibfnamefont {S.}~\bibnamefont
  {Maehrlein}}, \bibinfo {author} {\bibfnamefont {A.}~\bibnamefont {Paarmann}},
  \bibinfo {author} {\bibfnamefont {M.}~\bibnamefont {Wolf}}, \ and\ \bibinfo
  {author} {\bibfnamefont {T.}~\bibnamefont {Kampfrath}},\ }\bibfield  {title}
  {\enquote {\bibinfo {title} {Terahertz sum-frequency excitation of a
  raman-active phonon},}\ }\href@noop {} {\bibfield  {journal} {\bibinfo
  {journal} {Phys. Rev. Lett.}\ }\textbf {\bibinfo {volume} {119}},\ \bibinfo
  {pages} {127402} (\bibinfo {year} {2017}{\natexlab{b}})}\BibitemShut
  {NoStop}%
\bibitem [{\citenamefont {Gonze}\ and\ \citenamefont {Lee}(1997)}]{Gonze1997}%
  \BibitemOpen
  \bibfield  {author} {\bibinfo {author} {\bibfnamefont {X.}~\bibnamefont
  {Gonze}}\ and\ \bibinfo {author} {\bibfnamefont {C.}~\bibnamefont {Lee}},\
  }\bibfield  {title} {\enquote {\bibinfo {title} {{Dynamical matrices, Born
  effective charges, dielectric permittivity tensors, and interatomic force
  constants from density-functional perturbation theory}},}\ }\href@noop {}
  {\bibfield  {journal} {\bibinfo  {journal} {Phys. Rev. B}\ }\textbf {\bibinfo
  {volume} {55}},\ \bibinfo {pages} {10355} (\bibinfo {year}
  {1997})}\BibitemShut {NoStop}%
\bibitem [{\citenamefont {Juraschek}\ \emph {et~al.}(2017)\citenamefont
  {Juraschek}, \citenamefont {Fechner},\ and\ \citenamefont
  {Spaldin}}]{juraschek:2017}%
  \BibitemOpen
  \bibfield  {author} {\bibinfo {author} {\bibfnamefont {D.~M.}\ \bibnamefont
  {Juraschek}}, \bibinfo {author} {\bibfnamefont {M.}~\bibnamefont {Fechner}},
  \ and\ \bibinfo {author} {\bibfnamefont {N.~A.}\ \bibnamefont {Spaldin}},\
  }\bibfield  {title} {\enquote {\bibinfo {title} {{Ultrafast Structure
  Switching through Nonlinear Phononics}},}\ }\href@noop {} {\bibfield
  {journal} {\bibinfo  {journal} {Phys. Rev. Lett.}\ }\textbf {\bibinfo
  {volume} {118}},\ \bibinfo {pages} {054101} (\bibinfo {year}
  {2017})}\BibitemShut {NoStop}%
\bibitem [{\citenamefont {Dhar}\ \emph {et~al.}(1994)\citenamefont {Dhar},
  \citenamefont {Rogers},\ and\ \citenamefont {Nelson}}]{Dhar1994}%
  \BibitemOpen
  \bibfield  {author} {\bibinfo {author} {\bibfnamefont {L.}~\bibnamefont
  {Dhar}}, \bibinfo {author} {\bibfnamefont {J.~A.}\ \bibnamefont {Rogers}}, \
  and\ \bibinfo {author} {\bibfnamefont {K.~A.}\ \bibnamefont {Nelson}},\
  }\bibfield  {title} {\enquote {\bibinfo {title} {{Time-Resolved Vibrational
  Spectroscopy in the Impulsive Limit}},}\ }\href@noop {} {\bibfield  {journal}
  {\bibinfo  {journal} {Chem. Rev.}\ }\textbf {\bibinfo {volume} {94}},\
  \bibinfo {pages} {157--193} (\bibinfo {year} {1994})}\BibitemShut {NoStop}%
\bibitem [{\citenamefont {Subedi}(2015)}]{subedi:2015}%
  \BibitemOpen
  \bibfield  {author} {\bibinfo {author} {\bibfnamefont {A.}~\bibnamefont
  {Subedi}},\ }\bibfield  {title} {\enquote {\bibinfo {title} {Proposal for
  ultrafast switching of ferroelectrics using midinfrared pulses},}\
  }\href@noop {} {\bibfield  {journal} {\bibinfo  {journal} {Phys. Rev. B}\
  }\textbf {\bibinfo {volume} {92}},\ \bibinfo {pages} {214303} (\bibinfo
  {year} {2015})}\BibitemShut {NoStop}%
\bibitem [{\citenamefont {Fechner}\ and\ \citenamefont
  {Spaldin}(2016)}]{fechner:2016}%
  \BibitemOpen
  \bibfield  {author} {\bibinfo {author} {\bibfnamefont {M.}~\bibnamefont
  {Fechner}}\ and\ \bibinfo {author} {\bibfnamefont {N.~A.}\ \bibnamefont
  {Spaldin}},\ }\bibfield  {title} {\enquote {\bibinfo {title} {{Effects of
  intense optical phonon pumping on the structure and electronic properties of
  yttrium barium copper oxide}},}\ }\href@noop {} {\bibfield  {journal}
  {\bibinfo  {journal} {Phys. Rev. B}\ }\textbf {\bibinfo {volume} {94}},\
  \bibinfo {pages} {134307} (\bibinfo {year} {2016})}\BibitemShut {NoStop}%
\bibitem [{\citenamefont {Subedi}(2017)}]{subedi:2017}%
  \BibitemOpen
  \bibfield  {author} {\bibinfo {author} {\bibfnamefont {A.}~\bibnamefont
  {Subedi}},\ }\bibfield  {title} {\enquote {\bibinfo {title}
  {{Midinfrared-light-induced ferroelectricity in oxide paraelectrics via
  nonlinear phononics}},}\ }\href@noop {} {\bibfield  {journal} {\bibinfo
  {journal} {Phys. Rev. B}\ }\textbf {\bibinfo {volume} {95}},\ \bibinfo
  {pages} {134113} (\bibinfo {year} {2017})}\BibitemShut {NoStop}%
\bibitem [{\citenamefont {Mankowsky}\ \emph {et~al.}(2017)\citenamefont
  {Mankowsky}, \citenamefont {von Hoegen}, \citenamefont {F\"{o}rst},\ and\
  \citenamefont {Cavalleri}}]{Mankowsky_2:2017}%
  \BibitemOpen
  \bibfield  {author} {\bibinfo {author} {\bibfnamefont {R.}~\bibnamefont
  {Mankowsky}}, \bibinfo {author} {\bibfnamefont {A.}~\bibnamefont {von
  Hoegen}}, \bibinfo {author} {\bibfnamefont {M.}~\bibnamefont {F\"{o}rst}}, \
  and\ \bibinfo {author} {\bibfnamefont {A.}~\bibnamefont {Cavalleri}},\
  }\bibfield  {title} {\enquote {\bibinfo {title} {{Ultrafast Reversal of the
  Ferroelectric Polarization}},}\ }\href@noop {} {\bibfield  {journal}
  {\bibinfo  {journal} {Phys. Rev. Lett.}\ }\textbf {\bibinfo {volume} {118}},\
  \bibinfo {pages} {197601} (\bibinfo {year} {2017})}\BibitemShut {NoStop}%
\bibitem [{\citenamefont {Gu}\ and\ \citenamefont
  {Rondinelli}(2017{\natexlab{a}})}]{Gu2017}%
  \BibitemOpen
  \bibfield  {author} {\bibinfo {author} {\bibfnamefont {M.}~\bibnamefont
  {Gu}}\ and\ \bibinfo {author} {\bibfnamefont {J.~M.}\ \bibnamefont
  {Rondinelli}},\ }\bibfield  {title} {\enquote {\bibinfo {title} {{Role of
  orbital filling on nonlinear ionic Raman scattering in perovskite
  titanates}},}\ }\href@noop {} {\bibfield  {journal} {\bibinfo  {journal}
  {Phys. Rev. B}\ }\textbf {\bibinfo {volume} {95}},\ \bibinfo {pages} {024109}
  (\bibinfo {year} {2017}{\natexlab{a}})}\BibitemShut {NoStop}%
\bibitem [{\citenamefont {Gu}\ and\ \citenamefont
  {Rondinelli}(2017{\natexlab{b}})}]{Gu_2:2017}%
  \BibitemOpen
  \bibfield  {author} {\bibinfo {author} {\bibfnamefont {M.}~\bibnamefont
  {Gu}}\ and\ \bibinfo {author} {\bibfnamefont {J.~M.}\ \bibnamefont
  {Rondinelli}},\ }\bibfield  {title} {\enquote {\bibinfo {title} {{Nonlinear
  Phononic Control and Emergent Magnetism in Correlated Titanates}},}\ }\href
  {http://arxiv.org/abs/1710.00993} {\bibfield  {journal} {\bibinfo  {journal}
  {arXiv:1710.00993v1}\ } (\bibinfo {year} {2017}{\natexlab{b}})}\BibitemShut
  {NoStop}%
\bibitem [{\citenamefont {Fechner}\ \emph {et~al.}(2017)\citenamefont
  {Fechner}, \citenamefont {Sukhov}, \citenamefont {Chotorlishvili},
  \citenamefont {Kenel}, \citenamefont {Berakdar},\ and\ \citenamefont
  {Spaldin}}]{Fechner2017}%
  \BibitemOpen
  \bibfield  {author} {\bibinfo {author} {\bibfnamefont {M.}~\bibnamefont
  {Fechner}}, \bibinfo {author} {\bibfnamefont {A.}~\bibnamefont {Sukhov}},
  \bibinfo {author} {\bibfnamefont {L.}~\bibnamefont {Chotorlishvili}},
  \bibinfo {author} {\bibfnamefont {C.}~\bibnamefont {Kenel}}, \bibinfo
  {author} {\bibfnamefont {J.}~\bibnamefont {Berakdar}}, \ and\ \bibinfo
  {author} {\bibfnamefont {N.~A.}\ \bibnamefont {Spaldin}},\ }\bibfield
  {title} {\enquote {\bibinfo {title} {{Magnetophononics: ultrafast spin
  control through the lattice}},}\ }\href@noop {} {\bibfield  {journal}
  {\bibinfo  {journal} {arXiv:1707.03216v2}\ } (\bibinfo {year}
  {2017})}\BibitemShut {NoStop}%
\bibitem [{\citenamefont {Itin}\ and\ \citenamefont
  {Katsnelson}(2017)}]{Itin2017}%
  \BibitemOpen
  \bibfield  {author} {\bibinfo {author} {\bibfnamefont {A.~P.}\ \bibnamefont
  {Itin}}\ and\ \bibinfo {author} {\bibfnamefont {M~.I.}\ \bibnamefont
  {Katsnelson}},\ }\bibfield  {title} {\enquote {\bibinfo {title} {Efficient
  excitation of nonlinear phonons via chirped mid-infrared pulses: induced
  structural phase transitions},}\ }\href@noop {} {\bibfield  {journal}
  {\bibinfo  {journal} {arXiv:1707.02455v2}\ } (\bibinfo {year}
  {2017})}\BibitemShut {NoStop}%
\bibitem [{\citenamefont {Kresse}\ and\ \citenamefont
  {Furthm\"{u}ller}(1996{\natexlab{a}})}]{kresse:1996}%
  \BibitemOpen
  \bibfield  {author} {\bibinfo {author} {\bibfnamefont {G.}~\bibnamefont
  {Kresse}}\ and\ \bibinfo {author} {\bibfnamefont {J.}~\bibnamefont
  {Furthm\"{u}ller}},\ }\bibfield  {title} {\enquote {\bibinfo {title}
  {Efficiency of ab-initio total energy calculations for metals and
  semiconductors using a plane-wave basis set},}\ }\href@noop {} {\bibfield
  {journal} {\bibinfo  {journal} {Comput. Mat. Sci.}\ }\textbf {\bibinfo
  {volume} {6}},\ \bibinfo {pages} {15--50} (\bibinfo {year}
  {1996}{\natexlab{a}})}\BibitemShut {NoStop}%
\bibitem [{\citenamefont {Kresse}\ and\ \citenamefont
  {Furthm\"{u}ller}(1996{\natexlab{b}})}]{kresse2:1996}%
  \BibitemOpen
  \bibfield  {author} {\bibinfo {author} {\bibfnamefont {G.}~\bibnamefont
  {Kresse}}\ and\ \bibinfo {author} {\bibfnamefont {J.}~\bibnamefont
  {Furthm\"{u}ller}},\ }\bibfield  {title} {\enquote {\bibinfo {title}
  {Efficient iterative schemes for ab initio total-energy calculations using a
  plane-wave basis set},}\ }\href@noop {} {\bibfield  {journal} {\bibinfo
  {journal} {Phys. Rev. B}\ }\textbf {\bibinfo {volume} {54}},\ \bibinfo
  {pages} {11169} (\bibinfo {year} {1996}{\natexlab{b}})}\BibitemShut {NoStop}%
\bibitem [{\citenamefont {Togo}\ and\ \citenamefont {Tanaka}(2015)}]{phonopy}%
  \BibitemOpen
  \bibfield  {author} {\bibinfo {author} {\bibfnamefont {A.}~\bibnamefont
  {Togo}}\ and\ \bibinfo {author} {\bibfnamefont {I.}~\bibnamefont {Tanaka}},\
  }\bibfield  {title} {\enquote {\bibinfo {title} {First principles phonon
  calculations in materials science},}\ }\href@noop {} {\bibfield  {journal}
  {\bibinfo  {journal} {Scr. Mater.}\ }\textbf {\bibinfo {volume} {108}},\
  \bibinfo {pages} {1--5} (\bibinfo {year} {2015})}\BibitemShut {NoStop}%
\bibitem [{\citenamefont {Porezag}\ and\ \citenamefont
  {Pederson}(1996)}]{porezag:1996}%
  \BibitemOpen
  \bibfield  {author} {\bibinfo {author} {\bibfnamefont {D.}~\bibnamefont
  {Porezag}}\ and\ \bibinfo {author} {\bibfnamefont {M.~R.}\ \bibnamefont
  {Pederson}},\ }\bibfield  {title} {\enquote {\bibinfo {title} {{Infrared
  intensities and Raman-scattering activities within density-functional
  theory}},}\ }\href@noop {} {\bibfield  {journal} {\bibinfo  {journal} {Phys.
  Rev. B}\ }\textbf {\bibinfo {volume} {54}},\ \bibinfo {pages} {7830--7836}
  (\bibinfo {year} {1996})}\BibitemShut {NoStop}%
\bibitem [{\citenamefont {Monkhorst}\ and\ \citenamefont
  {Pack}(1976)}]{Monkhorst/Pack:1976}%
  \BibitemOpen
  \bibfield  {author} {\bibinfo {author} {\bibfnamefont {H.~J.}\ \bibnamefont
  {Monkhorst}}\ and\ \bibinfo {author} {\bibfnamefont {J.~D.}\ \bibnamefont
  {Pack}},\ }\bibfield  {title} {\enquote {\bibinfo {title} {Special points for
  brillouin-zone integrations},}\ }\href@noop {} {\bibfield  {journal}
  {\bibinfo  {journal} {Phys. Rev. B}\ }\textbf {\bibinfo {volume} {13}},\
  \bibinfo {pages} {5188--5192} (\bibinfo {year} {1976})}\BibitemShut {NoStop}%
\bibitem [{\citenamefont {Csonka}\ \emph {et~al.}(2009)\citenamefont {Csonka},
  \citenamefont {Perdew}, \citenamefont {Ruzsinszky}, \citenamefont
  {Philipsen}, \citenamefont {Leb\`{e}gue}, \citenamefont {Paier},
  \citenamefont {Vydrov},\ and\ \citenamefont {\'{A}ngy\'{a}n}}]{csonka:2009}%
  \BibitemOpen
  \bibfield  {author} {\bibinfo {author} {\bibfnamefont {G.~I.}\ \bibnamefont
  {Csonka}}, \bibinfo {author} {\bibfnamefont {J.~P.}\ \bibnamefont {Perdew}},
  \bibinfo {author} {\bibfnamefont {A.}~\bibnamefont {Ruzsinszky}}, \bibinfo
  {author} {\bibfnamefont {P.~H.~T.}\ \bibnamefont {Philipsen}}, \bibinfo
  {author} {\bibfnamefont {S.}~\bibnamefont {Leb\`{e}gue}}, \bibinfo {author}
  {\bibfnamefont {J.}~\bibnamefont {Paier}}, \bibinfo {author} {\bibfnamefont
  {O.~A.}\ \bibnamefont {Vydrov}}, \ and\ \bibinfo {author} {\bibfnamefont
  {J.~G.}\ \bibnamefont {\'{A}ngy\'{a}n}},\ }\bibfield  {title} {\enquote
  {\bibinfo {title} {Assessing the performance of recent density functionals
  for bulk solids},}\ }\href@noop {} {\bibfield  {journal} {\bibinfo  {journal}
  {Phys. Rev. B}\ }\textbf {\bibinfo {volume} {79}},\ \bibinfo {pages} {155107}
  (\bibinfo {year} {2009})}\BibitemShut {NoStop}%
\bibitem [{\citenamefont {Ederer}\ and\ \citenamefont
  {Spaldin}(2005)}]{Ederer/Spaldin:2005}%
  \BibitemOpen
  \bibfield  {author} {\bibinfo {author} {\bibfnamefont {C.}~\bibnamefont
  {Ederer}}\ and\ \bibinfo {author} {\bibfnamefont {N.~A.}\ \bibnamefont
  {Spaldin}},\ }\bibfield  {title} {\enquote {\bibinfo {title} {Weak
  ferromagnetism and magnetoelectric coupling in bismuth ferrite},}\
  }\href@noop {} {\bibfield  {journal} {\bibinfo  {journal} {Phys. Rev. B}\
  }\textbf {\bibinfo {volume} {71}},\ \bibinfo {pages} {060401(R)} (\bibinfo
  {year} {2005})}\BibitemShut {NoStop}%
\bibitem [{\citenamefont {Wei}\ \emph {et~al.}(2016)\citenamefont {Wei},
  \citenamefont {Fan}, \citenamefont {Wang}, \citenamefont {Liu}, \citenamefont
  {Zhang}, \citenamefont {Lv}, \citenamefont {Yang}, \citenamefont {Zhang},\
  and\ \citenamefont {Zhao}}]{Wei2016}%
  \BibitemOpen
  \bibfield  {author} {\bibinfo {author} {\bibfnamefont {L.}~\bibnamefont
  {Wei}}, \bibinfo {author} {\bibfnamefont {S.}~\bibnamefont {Fan}}, \bibinfo
  {author} {\bibfnamefont {X.~P.}\ \bibnamefont {Wang}}, \bibinfo {author}
  {\bibfnamefont {B.}~\bibnamefont {Liu}}, \bibinfo {author} {\bibfnamefont
  {Y.~Y.}\ \bibnamefont {Zhang}}, \bibinfo {author} {\bibfnamefont {X.~S.}\
  \bibnamefont {Lv}}, \bibinfo {author} {\bibfnamefont {Y.~G.}\ \bibnamefont
  {Yang}}, \bibinfo {author} {\bibfnamefont {H.~J.}\ \bibnamefont {Zhang}}, \
  and\ \bibinfo {author} {\bibfnamefont {X.}~\bibnamefont {Zhao}},\ }\bibfield
  {title} {\enquote {\bibinfo {title} {{Lattice dynamics of bismuth-deficient
  BiFeO$_3$ from first principles}},}\ }\href@noop {} {\bibfield  {journal}
  {\bibinfo  {journal} {Comput. Mater. Sci.}\ }\textbf {\bibinfo {volume}
  {111}},\ \bibinfo {pages} {374--379} (\bibinfo {year} {2016})}\BibitemShut
  {NoStop}%
\bibitem [{\citenamefont {Sosnowska}\ \emph {et~al.}(2002)\citenamefont
  {Sosnowska}, \citenamefont {Sch{\"a}fer}, \citenamefont {Kockelmann},
  \citenamefont {Andersen},\ and\ \citenamefont
  {Troyanchuk}}]{Sosnowska_et_al:2002}%
  \BibitemOpen
  \bibfield  {author} {\bibinfo {author} {\bibfnamefont {I.}~\bibnamefont
  {Sosnowska}}, \bibinfo {author} {\bibfnamefont {W.}~\bibnamefont
  {Sch{\"a}fer}}, \bibinfo {author} {\bibfnamefont {W.}~\bibnamefont
  {Kockelmann}}, \bibinfo {author} {\bibfnamefont {K.~H.}\ \bibnamefont
  {Andersen}}, \ and\ \bibinfo {author} {\bibfnamefont {I.~O.}\ \bibnamefont
  {Troyanchuk}},\ }\bibfield  {title} {\enquote {\bibinfo {title} {Crystal
  structure and spiral magnetic ordering in {BiFeO}$_3$ doped with
  manganese},}\ }\href@noop {} {\bibfield  {journal} {\bibinfo  {journal}
  {Appl. Phys. A}\ }\textbf {\bibinfo {volume} {74}},\ \bibinfo {pages}
  {S1040--S1042} (\bibinfo {year} {2002})}\BibitemShut {NoStop}%
\bibitem [{\citenamefont {Mildren}\ and\ \citenamefont
  {Rabeau}(2013)}]{Mildren2013}%
  \BibitemOpen
  \bibfield  {author} {\bibinfo {author} {\bibfnamefont {R.~P.}\ \bibnamefont
  {Mildren}}\ and\ \bibinfo {author} {\bibfnamefont {J.~R.}\ \bibnamefont
  {Rabeau}},\ }\href@noop {} {\emph {\bibinfo {title} {{Optical Engineering of
  Diamond}}}}\ (\bibinfo  {publisher} {Wiley},\ \bibinfo {year}
  {2013})\BibitemShut {NoStop}%
\bibitem [{\citenamefont {Ishioka}\ \emph {et~al.}(2016)\citenamefont
  {Ishioka}, \citenamefont {Hase}, \citenamefont {Masahiro},\ and\
  \citenamefont {Petek}}]{ishioka:2006}%
  \BibitemOpen
  \bibfield  {author} {\bibinfo {author} {\bibfnamefont {K.}~\bibnamefont
  {Ishioka}}, \bibinfo {author} {\bibfnamefont {M.}~\bibnamefont {Hase}},
  \bibinfo {author} {\bibfnamefont {K.}~\bibnamefont {Masahiro}}, \ and\
  \bibinfo {author} {\bibfnamefont {H.}~\bibnamefont {Petek}},\ }\bibfield
  {title} {\enquote {\bibinfo {title} {Coherent optical phonons in diamond},}\
  }\href@noop {} {\bibfield  {journal} {\bibinfo  {journal} {Appl. Phys.
  Lett.}\ }\textbf {\bibinfo {volume} {89}},\ \bibinfo {pages} {231916}
  (\bibinfo {year} {2016})}\BibitemShut {NoStop}%
\bibitem [{SUP()}]{SUPP-SFE:2017}%
  \BibitemOpen
  \href@noop {} {}\bibinfo {note} {See Supplemental Material at (Link) for
  illustrations of phase sensitivity, central-frequency dependence, and
  higher-order anharmonic couplings.}\BibitemShut {Stop}%
\bibitem [{\citenamefont {Shinohara}\ \emph
  {et~al.}(2010{\natexlab{a}})\citenamefont {Shinohara}, \citenamefont
  {Yabana}, \citenamefont {Kawashita}, \citenamefont {Iwata}, \citenamefont
  {Otobe},\ and\ \citenamefont {Bertsch}}]{Shinohara2010}%
  \BibitemOpen
  \bibfield  {author} {\bibinfo {author} {\bibfnamefont {Y.}~\bibnamefont
  {Shinohara}}, \bibinfo {author} {\bibfnamefont {K.}~\bibnamefont {Yabana}},
  \bibinfo {author} {\bibfnamefont {Y.}~\bibnamefont {Kawashita}}, \bibinfo
  {author} {\bibfnamefont {J.-I.}\ \bibnamefont {Iwata}}, \bibinfo {author}
  {\bibfnamefont {T.}~\bibnamefont {Otobe}}, \ and\ \bibinfo {author}
  {\bibfnamefont {G.~F.}\ \bibnamefont {Bertsch}},\ }\bibfield  {title}
  {\enquote {\bibinfo {title} {{Coherent phonon generation in time-dependent
  density functional theory}},}\ }\href@noop {} {\bibfield  {journal} {\bibinfo
   {journal} {Phys. Rev. B}\ }\textbf {\bibinfo {volume} {82}},\ \bibinfo
  {pages} {155110} (\bibinfo {year} {2010}{\natexlab{a}})}\BibitemShut
  {NoStop}%
\bibitem [{\citenamefont {Shinohara}\ \emph
  {et~al.}(2010{\natexlab{b}})\citenamefont {Shinohara}, \citenamefont
  {Kawashita}, \citenamefont {Iwata}, \citenamefont {Yabana}, \citenamefont
  {Otobe},\ and\ \citenamefont {Bertsch}}]{Shinohara2:2010}%
  \BibitemOpen
  \bibfield  {author} {\bibinfo {author} {\bibfnamefont {Y.}~\bibnamefont
  {Shinohara}}, \bibinfo {author} {\bibfnamefont {Y.}~\bibnamefont
  {Kawashita}}, \bibinfo {author} {\bibfnamefont {J.-I.}\ \bibnamefont
  {Iwata}}, \bibinfo {author} {\bibfnamefont {K.}~\bibnamefont {Yabana}},
  \bibinfo {author} {\bibfnamefont {T.}~\bibnamefont {Otobe}}, \ and\ \bibinfo
  {author} {\bibfnamefont {G.~F.}\ \bibnamefont {Bertsch}},\ }\bibfield
  {title} {\enquote {\bibinfo {title} {{First-principles description for
  coherent phonon generation in diamond}},}\ }\href@noop {} {\bibfield
  {journal} {\bibinfo  {journal} {J.~Phys.: Condens. Matter}\ }\textbf
  {\bibinfo {volume} {22}},\ \bibinfo {pages} {384212} (\bibinfo {year}
  {2010}{\natexlab{b}})}\BibitemShut {NoStop}%
\bibitem [{\citenamefont {Green}\ \emph {et~al.}(2016)\citenamefont {Green},
  \citenamefont {Kovalev}, \citenamefont {Asgekar}, \citenamefont {Geloni},
  \citenamefont {Lehnert}, \citenamefont {Golz}, \citenamefont {Kuntzsch},
  \citenamefont {Bauer}, \citenamefont {Hauser}, \citenamefont {Voigtlaender},
  \citenamefont {Wustmann}, \citenamefont {Koesterke}, \citenamefont {Schwarz},
  \citenamefont {Freitag}, \citenamefont {Arnold}, \citenamefont {Teichert},
  \citenamefont {Justus}, \citenamefont {Seidel}, \citenamefont {Ilgner},
  \citenamefont {Awari}, \citenamefont {Nicoletti}, \citenamefont {Kaiser},
  \citenamefont {Laplace}, \citenamefont {Rajasekaran}, \citenamefont {Zhang},
  \citenamefont {Winnerl}, \citenamefont {Schneider}, \citenamefont {Schay},
  \citenamefont {Lorincz}, \citenamefont {Rauscher}, \citenamefont {Radu},
  \citenamefont {Maehrlein}, \citenamefont {Kim}, \citenamefont {Lee},
  \citenamefont {Kampfrath}, \citenamefont {Wall}, \citenamefont {Heberle},
  \citenamefont {Malnasi-Csizmadia}, \citenamefont {Steiger}, \citenamefont
  {M\"{u}ller}, \citenamefont {Helm}, \citenamefont {Schramm}, \citenamefont
  {Cowan}, \citenamefont {Michel}, \citenamefont {Cavalleri}, \citenamefont
  {Fisher}, \citenamefont {Stojanovic},\ and\ \citenamefont
  {Gensch}}]{Green2016}%
  \BibitemOpen
  \bibfield  {author} {\bibinfo {author} {\bibfnamefont {B.}~\bibnamefont
  {Green}}, \bibinfo {author} {\bibfnamefont {S.}~\bibnamefont {Kovalev}},
  \bibinfo {author} {\bibfnamefont {V.}~\bibnamefont {Asgekar}}, \bibinfo
  {author} {\bibfnamefont {G.}~\bibnamefont {Geloni}}, \bibinfo {author}
  {\bibfnamefont {U.}~\bibnamefont {Lehnert}}, \bibinfo {author} {\bibfnamefont
  {T.}~\bibnamefont {Golz}}, \bibinfo {author} {\bibfnamefont {M.}~\bibnamefont
  {Kuntzsch}}, \bibinfo {author} {\bibfnamefont {C.}~\bibnamefont {Bauer}},
  \bibinfo {author} {\bibfnamefont {J.}~\bibnamefont {Hauser}}, \bibinfo
  {author} {\bibfnamefont {J.}~\bibnamefont {Voigtlaender}}, \bibinfo {author}
  {\bibfnamefont {B.}~\bibnamefont {Wustmann}}, \bibinfo {author}
  {\bibfnamefont {I.}~\bibnamefont {Koesterke}}, \bibinfo {author}
  {\bibfnamefont {M.}~\bibnamefont {Schwarz}}, \bibinfo {author} {\bibfnamefont
  {M.}~\bibnamefont {Freitag}}, \bibinfo {author} {\bibfnamefont
  {A.}~\bibnamefont {Arnold}}, \bibinfo {author} {\bibfnamefont
  {J.}~\bibnamefont {Teichert}}, \bibinfo {author} {\bibfnamefont
  {M.}~\bibnamefont {Justus}}, \bibinfo {author} {\bibfnamefont
  {W.}~\bibnamefont {Seidel}}, \bibinfo {author} {\bibfnamefont
  {C.}~\bibnamefont {Ilgner}}, \bibinfo {author} {\bibfnamefont
  {N.}~\bibnamefont {Awari}}, \bibinfo {author} {\bibfnamefont
  {D.}~\bibnamefont {Nicoletti}}, \bibinfo {author} {\bibfnamefont
  {S.}~\bibnamefont {Kaiser}}, \bibinfo {author} {\bibfnamefont
  {Y.}~\bibnamefont {Laplace}}, \bibinfo {author} {\bibfnamefont
  {S.}~\bibnamefont {Rajasekaran}}, \bibinfo {author} {\bibfnamefont
  {L.}~\bibnamefont {Zhang}}, \bibinfo {author} {\bibfnamefont
  {S.}~\bibnamefont {Winnerl}}, \bibinfo {author} {\bibfnamefont
  {H.}~\bibnamefont {Schneider}}, \bibinfo {author} {\bibfnamefont
  {G.}~\bibnamefont {Schay}}, \bibinfo {author} {\bibfnamefont
  {I.}~\bibnamefont {Lorincz}}, \bibinfo {author} {\bibfnamefont {A.~A.}\
  \bibnamefont {Rauscher}}, \bibinfo {author} {\bibfnamefont {I.}~\bibnamefont
  {Radu}}, \bibinfo {author} {\bibfnamefont {S.}~\bibnamefont {Maehrlein}},
  \bibinfo {author} {\bibfnamefont {T.~H.}\ \bibnamefont {Kim}}, \bibinfo
  {author} {\bibfnamefont {J.~S.}\ \bibnamefont {Lee}}, \bibinfo {author}
  {\bibfnamefont {T.}~\bibnamefont {Kampfrath}}, \bibinfo {author}
  {\bibfnamefont {S.}~\bibnamefont {Wall}}, \bibinfo {author} {\bibfnamefont
  {J.}~\bibnamefont {Heberle}}, \bibinfo {author} {\bibfnamefont
  {A.}~\bibnamefont {Malnasi-Csizmadia}}, \bibinfo {author} {\bibfnamefont
  {A.}~\bibnamefont {Steiger}}, \bibinfo {author} {\bibfnamefont {A.~S.}\
  \bibnamefont {M\"{u}ller}}, \bibinfo {author} {\bibfnamefont
  {M.}~\bibnamefont {Helm}}, \bibinfo {author} {\bibfnamefont {U.}~\bibnamefont
  {Schramm}}, \bibinfo {author} {\bibfnamefont {T.}~\bibnamefont {Cowan}},
  \bibinfo {author} {\bibfnamefont {P.}~\bibnamefont {Michel}}, \bibinfo
  {author} {\bibfnamefont {A.}~\bibnamefont {Cavalleri}}, \bibinfo {author}
  {\bibfnamefont {A.~S.}\ \bibnamefont {Fisher}}, \bibinfo {author}
  {\bibfnamefont {N.}~\bibnamefont {Stojanovic}}, \ and\ \bibinfo {author}
  {\bibfnamefont {M.}~\bibnamefont {Gensch}},\ }\bibfield  {title} {\enquote
  {\bibinfo {title} {{High-Field High-Repetition-Rate Sources for the Coherent
  THz Control of Matter}},}\ }\href@noop {} {\bibfield  {journal} {\bibinfo
  {journal} {Scientific Reports}\ }\textbf {\bibinfo {volume} {6}},\ \bibinfo
  {pages} {22256} (\bibinfo {year} {2016})}\BibitemShut {NoStop}%
\bibitem [{\citenamefont {Gruene}\ \emph {et~al.}(2008)\citenamefont {Gruene},
  \citenamefont {Rayner}, \citenamefont {Redlich}, \citenamefont {van~der
  Meer}, \citenamefont {Lyon}, \citenamefont {Meijer},\ and\ \citenamefont
  {Fielicke}}]{Gruene2008}%
  \BibitemOpen
  \bibfield  {author} {\bibinfo {author} {\bibfnamefont {P.}~\bibnamefont
  {Gruene}}, \bibinfo {author} {\bibfnamefont {D.~M.}\ \bibnamefont {Rayner}},
  \bibinfo {author} {\bibfnamefont {B.}~\bibnamefont {Redlich}}, \bibinfo
  {author} {\bibfnamefont {A.~F.~G.}\ \bibnamefont {van~der Meer}}, \bibinfo
  {author} {\bibfnamefont {J.~T.}\ \bibnamefont {Lyon}}, \bibinfo {author}
  {\bibfnamefont {G.}~\bibnamefont {Meijer}}, \ and\ \bibinfo {author}
  {\bibfnamefont {A.}~\bibnamefont {Fielicke}},\ }\bibfield  {title} {\enquote
  {\bibinfo {title} {Structures of neutral {Au}$_7$, {Au}$_{19}$, and
  {Au}$_{20}$ clusters in the gas phase},}\ }\href
  {http://science.sciencemag.org/content/321/5889/674.abstract} {\bibfield
  {journal} {\bibinfo  {journal} {Science}\ }\textbf {\bibinfo {volume}
  {321}},\ \bibinfo {pages} {674} (\bibinfo {year} {2008})}\BibitemShut
  {NoStop}%
\bibitem [{\citenamefont {Liu}\ \emph {et~al.}(2017)\citenamefont {Liu},
  \citenamefont {Bromberger}, \citenamefont {Cartella}, \citenamefont {Gebert},
  \citenamefont {F\"{o}rst},\ and\ \citenamefont {Cavalleri}}]{Liu2017}%
  \BibitemOpen
  \bibfield  {author} {\bibinfo {author} {\bibfnamefont {B.}~\bibnamefont
  {Liu}}, \bibinfo {author} {\bibfnamefont {H.}~\bibnamefont {Bromberger}},
  \bibinfo {author} {\bibfnamefont {A.}~\bibnamefont {Cartella}}, \bibinfo
  {author} {\bibfnamefont {T.}~\bibnamefont {Gebert}}, \bibinfo {author}
  {\bibfnamefont {M.}~\bibnamefont {F\"{o}rst}}, \ and\ \bibinfo {author}
  {\bibfnamefont {A.}~\bibnamefont {Cavalleri}},\ }\bibfield  {title} {\enquote
  {\bibinfo {title} {{Generation of narrowband, high-intensity,
  carrier-envelope phase-stable pulses tunable between 4 and 18 THz}},}\
  }\href@noop {} {\bibfield  {journal} {\bibinfo  {journal} {Opt. Lett.}\
  }\textbf {\bibinfo {volume} {42}},\ \bibinfo {pages} {129--131} (\bibinfo
  {year} {2017})}\BibitemShut {NoStop}%
\end{thebibliography}

%

\end{document}